\documentclass[preprint,superscriptaddress,floatfix,preprintnumbers,nofootinbib,altaffilletter]{revtex4-1}
\usepackage{graphicx}
\usepackage{subfigure}
\usepackage{dcolumn}
\usepackage{bm}
\usepackage{amssymb}
\usepackage{xcolor}
\usepackage[utf8]{inputenc}
\usepackage[OT1]{fontenc}
\usepackage{yfonts,amsmath,amsthm,amsfonts,amssymb,amscd, ulem}
\usepackage{enumerate}
\usepackage{fancyhdr}
\usepackage{mathtools}
\usepackage{mathrsfs}
\usepackage{cancel}
\usepackage{slashed}
\usepackage[flushleft]{threeparttable}
\usepackage[colorlinks=true, citecolor=purple, linkcolor=blue]{hyperref}
\usepackage{url}
\usepackage{makecell,booktabs}
\usepackage{braket}
\usepackage{relsize}
\usepackage{multirow}
\usepackage{verbatim}
\usepackage{txfonts}
\usepackage{upgreek}
\usepackage{extarrows}
\usepackage{array}
\usepackage[T1]{fontenc}

\begin{document}
\title{Long-lived Searches of Vector-like Lepton and Its Accompanying Scalar at Colliders}

\author{Qing-Hong Cao}
\email{qinghongcao@pku.edu.cn}
\affiliation{School of Physics, Peking University, Beijing 100871, China}
\affiliation{Center for High Energy Physics, Peking University, Beijing 100871, China}

\author{Jinhui Guo}
\email{guojh23@pku.edu.cn}
\affiliation{School of Physics and State Key Laboratory of Nuclear Physics and Technology, Peking University, Beijing 100871, China}

\author{Jia Liu}
\email{jialiu@pku.edu.cn}
\affiliation{School of Physics and State Key Laboratory of Nuclear Physics and Technology, Peking University, Beijing 100871, China}
\affiliation{Center for High Energy Physics, Peking University, Beijing 100871, China}

\author{Yan Luo}
\email{ly23@stu.pku.edu.cn}
\affiliation{School of Physics and State Key Laboratory of Nuclear Physics and Technology, Peking University, Beijing 100871, China}

\author{Xiao-Ping Wang}
\email{hcwangxiaoping@buaa.edu.cn}
\affiliation{School of Physics, Beihang University, Beijing 100083, China}
\affiliation{Beijing Key Laboratory of Advanced Nuclear Materials and Physics, Beihang University, Beijing 100191, China}

\begin{abstract}
Recently, the vector-like leptons (VLLs) as a simple extension to the standard model (SM) have attracted widespread attention both in theory and experiments. The present collider searches mainly focus on the studies of their prompt decays, which prefer a relatively large coupling. In this paper, we concentrate on searches for long-lived signatures of the singlet VLLs $F$ or their accompanying scalar particles $\phi$ both in the hadronic and electronic colliders. The long-lived signatures are naturally induced from small chiral mass mixing between VLLs and SM leptons. Two specific models distinguished by whether the VLLs couple to scalar particles are introduced to realize the aforementioned features. For long-lived VLLs case, we find that with the kink track method the sensitivities at future HL-LHC with $\sqrt{s}=14~\text{TeV}$ can reach the regions for VLL mass $m_F \in [200,1100]~\text{GeV}$ and the mass mixing parameter $\theta_L \in [10^{-10},3\times 10^{-8}]$. For the long-lived accompanying scalar particle case, by fixing VLLs or scalar mass, or the mass ratio between VLL and the accompanying scalar, we explore the projected sensitivities through the time delay and displaced vertex strategies, which can probe the regions for $m_F \in [200,1200]~\text{GeV}$ and coupling $y\theta_L\in [10^{-11},10^{-6}]$. Furthermore, we also explore the long-lived accompanying scalars at the future CEPC provided that the VLLs can couple to the SM first-generation leptons. We find that CEPC has good performances for $m_\phi < 120~\text{GeV}$ and $m_F<1200~\text{GeV}$. These long-lived searches are complementary to previous studies, which opens the door towards the smaller coupling regions.

\end{abstract}

\maketitle
\tableofcontents

\section{Introduction}
\label{sec:int}
Vector-like fermion (VLF) stands as one of the simplest extensions of the Standard Model (SM), transforming as a non-chiral representation of SM gauge group and potentially originating from scenarios like string-theory models or grand unified theories (GUTs), beyond the three-generation fermions~\cite{Frampton:1999xi}.
Heavy VLFs can realize the flavor democracy~\cite{Baspehlivan:2022qet}, and there have been many searches for vector-like quarks for decades, both at the Tevatron~\cite{D0:2010ckq,CDF:2009gat} and the Large Hadron Collider (LHC)~\cite{CMS:2011lcm,ATLAS:2017vdo,ATLAS:2018ziw,CMS:2019eqb,CMS:2022fck}. If VLFs only couple to SM leptons, they are usually referred to as vector-like leptons (VLLs) and have been paid more attention recently. They are often classified by their coupled SM lepton generation, where they possess the same quantum numbers as the SM leptons. 

Conventional searches for these particles usually focus on the associated production or pair production followed by the prompt decays of VLLs, such as the searches for VLL doublet coupled to the third-generation, conducted by ATLAS and CMS \cite{ATLAS:2023sbu, CMS:2019hsm, CMS:2022nty,Muse:2022xgu}. Collider phenomenology research on different decay modes of VLLs has been conducted \cite{Thomas:1998wy, Sher:1995tc, Frampton:1999xi, Falkowski:2013jya, Bhattiprolu:2019vdu,Guedes:2021oqx}. Additionally, a dedicated study explores a model featuring a VLL doublet and scalars, capable of generating the requisite baryon asymmetry~\cite{Bell:2019mbn}. Moreover, investigations have explored the searches for $SU(2)_L$ singlet VLLs with a small mass mixing to SM leptons, originating from the Yukawa coupling to the Higgs field. These searches have been studied at the LHC and proposed for future $e^-e^+/\mu^+\mu^-$ colliders~\cite{Dermisek:2014qca, Kumar:2015tna, Shang:2021mgn, Guo:2023jkz}. Furthermore, other noteworthy discussions regarding VLFs in the context of Higgs physics and new physics have been explored in Refs. \cite{Kearney:2012zi,Halverson:2014nwa,Arina:2012aj,Schwaller:2013hqa,Dermisek:2013gta,Ishiwata:2013gma,Dermisek:2014cia,Dermisek:2015oja,Chen:2016lsr,Poh:2017tfo,Xu:2018pnq,Zheng:2019kqu,Freitas:2020ttd,Bissmann:2020lge,Dermisek:2021ajd,Kawamura:2021ygg,Yang:2021dtc,Li:2022vsg,Raju:2022zlv,Li:2022hzl,Shang:2023rfv,Kawamura:2023zuo,Bhattiprolu:2023yxa,Ghosh:2023dhj,Saez:2021qta,Crivellin:2018qmi,Crivellin:2021bkd,Crivellin:2021rbq,Kim:2018mks,Alhazmi:2018whk}.

It is important to highlight that previous VLL searches have predominantly focused on VLL production and subsequent prompt decays, emphasizing strong couplings during investigations. In contrast, recent research endeavors have explored alternative aspects of VLLs and their interactions. For instance, one study has examined the interplay between an inert Higgs doublet dark matter and a VLL triplet, specifically considering potential displaced vertices signals detectable at the MATHUSLA detector~\cite{Bandyopadhyay:2023joz}. Furthermore, another recent investigation delved into a model featuring a weak-singlet VLL denoted as $\tau'$ and a complex scalar, studying the phenomenology of long-lived pseudoscalar signals at the LHC~\cite{Bernreuther:2023uxh}. In their model, the complex scalar possesses a non-zero vacuum expectation value (vev), resulting in the pseudoscalar mass being suppressed to a few GeV. In contrast, in our model, the scalar mass can reach several hundred GeV. Furthermore, our research primarily focuses on cases where VLLs couple with either electrons or muons, and we explore scenarios involving long-lived weak-singlet VLLs.

In our study, we focus on VLLs that share the same charge as right-handed fermions in the SM. We find that it's natural for these VLLs and their accompanying scalar particles to exhibit long-lived signatures. These long-lived features, resulting from left-handed mixing, naturally involve tiny couplings, which are also invaluable in addressing issues related to lepton flavor violations (LFV). To illustrate this, we present two models featuring an $SU(2)_L$ singlet charged VLL, denoted as $F^\pm$, which undergoes mixing with SM leptons via direct mass mixing. We propose a search strategy for these novel particles based on their long-lived signatures. In the first model, we seek long-lived VLLs $F^\pm$ with a kink track signature, where the production of $F^-F^+$ pairs occurs through the Drell-Yan process, followed by their long-lived subsequent decays to $Z\ell^\pm$, $W^\pm \nu_\ell$ or $h\ell^\pm$. In the second model, we introduce a long-lived accompanying scalar particle coupled to both $F^\pm$ and SM leptons via a substantial Yukawa interaction, and we explore its detection using time delay and displaced vertex signatures. Our first chosen search platform is the high-luminosity LHC (HL-LHC), a traditional facility for exploring heavy new particles. These searches extend the parameter space set by ATLAS and CMS, opening new possibilities for discovery.

Moreover, in contrast to the first model, the second model, despite its heavier mass, holds exploration potential through the $t$ channel production at future $e^-e^+$ colliders, especially when VLLs interact with electrons. To exemplify this, we consider the Circular Electron Positron Collider (CEPC) as an illustrative case. By examining the displaced vertex signatures associated with the scalar particles, CEPC exhibits excellent sensitivity, particularly in lighter mass regions, thanks to its pristine experimental conditions and lower center-of-mass energy.

We organize the paper as follows. In section~\ref{sec:model}, we describe the $SU(2)_L$ singlet VLL models with singlet scalar or without scalar, respectively, and their possible decay channels. In section~\ref{sec:excon}, we discuss the existing constraints from collider searches, LFV processes, and lepton $g-2$ measurements. In section~\ref{sec:LLP}, we discuss the long-lived particle (LLP) signatures and their detection at the HL-LHC and CEPC. In section~\ref{sec:conclusions}, we conclude.

\section{The Models}
\label{sec:model}
We expand SM with an extra $SU(2)_L$ singlet VLL $F^\pm$ and a real scalar singlet $\phi$, and their SM gauge group property are listed in Tab. \ref{tab:1}.
\begin{table}[ht]
\centering
\begin{tabular}{c|c|c}
\hline
Gauge  Group                   & Fermion Field ($F^\pm$)  & Scalar Fields ($\phi$)        
\\ \hline
$SU(2)_L\times U(1)_Y$    & (1,$-2$)   & (1,0)   \\
\hline
\end{tabular}%
\caption{Gauge charges of new particles and relevant SM particles in SM groups.}
\label{tab:1}
\end{table}

In the following subsections, we discuss two scenarios: one with only the VLL $F^\pm$ and another with both the VLL $F^\pm$ and the scalar $\phi$. Additionally, we focus on cases where VLL interacts exclusively with a single lepton generation, such as the muon or electron.

\subsection{Vector-like Lepton Scenario (VLLS)}\label{subsec:VLLS}

\begin{figure}[htbp]
    \centering
    \includegraphics[width=0.5\linewidth]{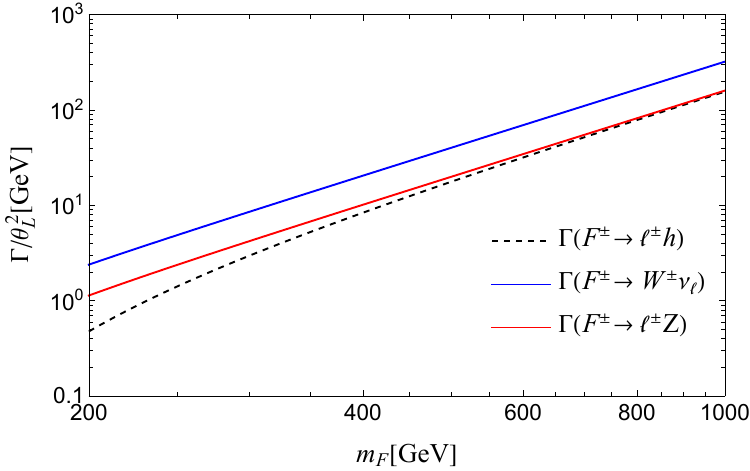}
    \caption{The decay widths of the main decay channels of $F^\pm$, rescaled by $1/\theta_L^2$, are depicted in various colors.}
    \label{fig:decay-width-F}
\end{figure}

The effective Lagrangian related to $F^\pm$ reads as
\begin{equation}\label{eq:1}
    \mathcal{L}_{\rm eff}^{F}\supset \bar{F}^0iD_\mu\gamma^\mu F^0 +\bar{L}^0 iD_\mu\gamma^\mu L^0 + \bar{\ell}^0_R i D_\mu \gamma^\mu \ell^0_R - m^0_F\bar{ F}^0 F^0 - \left(m^0_\ell + \frac{m^0_\ell}{v_h}h\right) \bar{\ell}^0 \ell^0-(\delta \bar {F}^0_L \ell^0_R + {\rm h.c.}),
  \end{equation}
where the covariant derivative $D_\mu = \partial_\mu - ig' \frac{Y}{2}B_\mu-igT^iW_\mu^i$, $B_\mu$ and $W_\mu^i$ represent the gauge fields for the SM $U(1)_Y$ and $SU(2)_L$ groups, respectively. The constants $g'$ and $g$ are their associated coupling constants. The lepton receives its mass $m^0_\ell$ from Higgs mechanism, $v_h$ is the Higgs vacuum and $h$ is the Higgs. Furthermore, $\ell$ denotes charged lepton singlet, and $L$ represents weak isospin doublets. All fields with a superscript `0' refer to interaction-eigenstates. 
Although the Higgs boson could potentially have a Yukawa coupling to SM leptons in the form of $y'\bar{L}^0 H F_R^0$, we deliberately disregard this possibility by setting $y'=0$, focusing on the long-lived signature of $F$.
The mass of the VLL is primarily determined by its mass term, with a small correction introduced by the $\delta$ term.
The mass part of Eq. (\ref{eq:1}) can be written as
\begin{equation} \label{eq:2}
   \mathcal{L}_{\rm mass}^{F}= \left(\bar{F}^0_L, \bar{\ell}^0_L\right)\left(\begin{matrix}m^0_F & \delta\\ 0 & m^0_\ell \end{matrix}\right)\left(\begin{matrix}F^0_R \\ \ell^0_R \end{matrix}\right)+{\rm h.c.}=\left(\bar{F}^0_L, \bar{\ell}^0_L\right)M\left(\begin{matrix}F^0_R \\ \ell^0_R \end{matrix}\right)+{\rm h.c.}=\left(\bar{F}_L, \bar{\ell}_L\right) U_L M U^\dagger_R\left(\begin{matrix}F_R \\ \ell_R \end{matrix}\right)+{\rm h.c.},
\end{equation}
where the unitary matrices are introduced to diagonalize the mass matrix
\begin{equation}\label{eq:3}
    U_L M U^\dagger_R=\text{diag} \left(m_F,m_\ell\right), 
\end{equation}
 and obtain the mass eigenstates $F$ and $\ell$. While the unitary matrices can be parameterized as
\begin{equation}\label{eq:4}
    U_L=\left(\begin{matrix}\cos{\theta_L} & -\sin{\theta_L}\\ \sin{\theta_L} & \cos\theta_L \end{matrix} \right), U_R=\left(\begin{matrix}\cos{\theta_R} & -\sin{\theta_R}\\ \sin{\theta_R} & \cos\theta_R \end{matrix} \right),
\end{equation}
with
\begin{equation}\label{eq:5}
\begin{aligned}
    \tan\theta_R &=-\frac{2m^0_F\delta}{(m^0_F)^2-(m^0_\ell)^2-\delta^2+\sqrt{((m^0_F)^2-(m^0_\ell)^2+\delta^2)^2+4(m^0_\ell)^2\delta^2}}\simeq -\frac{\delta}{m^0_F}, \\
    \tan\theta_L &=-\frac{2m ^0_\ell\delta}{(m^0_F)^2-(m^0_\ell)^2+\delta^2+\sqrt{((m^0_F)^2-(m^0_\ell)^2+\delta^2)^2+4(m^0_\ell)^2\delta^2}}\simeq -\frac{m^0_\ell\delta}{(m^0_F)^2}\simeq\frac{m^0_{\ell}}{m_F^0} \tan\theta_R ,\\
\end{aligned}    
\end{equation}
The approximation is valid in the limits where $m_F^0 \gg m_\ell^0, ~\delta$. Due to this hierarchy, the mixing angle $\theta_L$ is significantly suppressed by an additional factor of $m_\ell^0 / m_F^0$ in comparison to $\theta_R$. This is a natural outcome as we introduce a heavy VLL, with the same gauge charge as the right-handed SM fermion. Thus, its mixing with right-handed fermions is considerably more straightforward than with the left-handed counterparts.
Consequently, we can establish $|\theta_L|\ll|\theta_R|\ll 1$. We also can simplify the mass eigenvalues
\begin{align}
 m_F&\simeq m^0_F+\frac{\delta^2}{2 m^0_F} \simeq m^0_F
 \label{mF-mass}\\
 m_\ell &\simeq m^0_\ell\left(1- \frac{1}{2}\left(\frac{\delta}{m^0_F}\right)^2 \right).
 \label{ml-mass}
\end{align}
To the lowest order of $\theta_L(\theta_R)$, the Lagrangian (\ref{eq:1}) in the mass eigenstates can be expressed as
\begin{equation}
\begin{aligned}\label{eq:7}
    \mathcal{L}_{\rm eff}^F\supset ~& i\bar{F}\gamma^\mu\partial_\mu F+i\bar{\ell}\gamma^\mu\partial_\mu \ell- m_F\bar{F} F - m_\ell \bar{\ell} \ell\\
    & -\frac{g}{2}\left( W^3_{\mu}\left(\bar{\ell}_L\gamma^\mu \ell_L-\theta_L\bar{F}_L\gamma^\mu \ell_L -\theta_L\bar{\ell}_L\gamma^\mu F_L\right) + \left(\left(W^1_{\mu}-iW^2_{\mu}\right)\left(-\theta_L \bar\nu_L\gamma^\mu F_L+\bar\nu_L\gamma^\mu \ell_L\right)+{\rm h.c.} \right)\right)\\
    & -g'B_\mu\Big(\bar F\gamma^\mu F + \frac{1}{2}\bar \ell_L\gamma^\mu \ell_L + \bar \ell_R\gamma^\mu \ell_R +\frac{\theta_L}{2}\bar{F}_L\gamma^\mu \ell_L + \frac{\theta_L}{2}\bar{\ell}_L\gamma^\mu F_L\Big) \\
    & -\frac{m_\ell}{v_h}h\left(\bar{\ell}_L \ell_R + \theta_L\theta_R \bar{F}_L F_R -\theta_L \bar{F}_L \ell_R - \theta_R \bar{\ell}_L F_R +\rm{h.c.} \right)\\
    \supset ~&\bar{F} \left(i\partial_\mu-e A_\mu+e\tan\theta_W Z_\mu\right)\gamma^\mu F 
    - m_F\bar{F}F  - m_\ell \bar{\ell} \ell -\frac{m_\ell}{v_h}h \bar{\ell} \ell \\
    & +\frac{1}{2}\frac{e}{\sin{\theta_W}\cos{\theta_W}}\theta_L Z_\mu \left(\bar{F}_L\gamma^\mu \ell_L +{\rm h.c.}\right) -\frac{e}{\sqrt{2}\sin\theta_W} \theta_L \left(W_\mu^+ \bar{\nu}_L\gamma^\mu F_L +{\rm h.c.}\right)\\
    & + \frac{m_\ell}{v_h}h \left(\theta_L \bar{F}_L \ell_R + \theta_R \bar{\ell}_L F_R +\rm{h.c.} \right),
\end{aligned}
\end{equation}
where $\theta_W$ is the Weinberg angle. In the weak interactions, exclusively the left-handed fermions engage, as Eq. (\ref{eq:1}). Consequently, within the fermion ($F$) interaction with SM leptons mediated by gauge bosons, solely the left-handed components ($F_L$) are active, characterized by couplings proportional to $\theta_L$. 
In the context of Higgs interactions according to Eq. (\ref{eq:1}), both left-handed and right-handed interactions occur. The term with a coupling proportional to $\theta_R$ predominates due to $\theta_L\simeq \frac{m^0_\ell}{m^0_F}\theta_R\ll \theta_R$, as depicted in the final line in Eq. (\ref{eq:7}). When analyzing the decay of $F^\pm\to h \ell^\pm$, we focus solely on the dominant contribution proportional to $\theta_R$. For $m_F>m_W, m_Z$ and $m_h$, the corresponding decay width of $F$ can be deduced to
\begin{align}\label{eq:FtoW}
    &\Gamma(F^\pm \to \nu_\ell W^\pm)=\frac{\theta_L^2 g^2}{64 \pi} \frac{(m_F^2-m_W^2)^2(m_F^2+2m_W^2)}{m_F^3 m_W^2}, \\
    &\Gamma(F^\pm \to \ell^\pm Z)\simeq \frac{\theta_L^2 g_Z^2}{128 \pi} \frac{(m_F^2-m_Z^2)^2(m_F^2+2m_Z^2)}{m_F^3 m_Z^2}, \\
    &\Gamma(F^\pm\to h \ell^\pm)\simeq \frac{\theta_R^2 m_\ell^2}{ m_F^2}\frac{1}{32\pi}
    \frac{(m_F^2-m_h^2)^2}{m_F v_h^2} = \frac{\theta_L^2 g^2}{128\pi}
    \frac{(m_F^2-m_h^2)^2}{m_F m_W^2},
\label{eq:FtoZ}
\end{align}
where $g = e/\sin{\theta_W} = 2m_W/v_h$, $g_Z = e/(\sin{\theta_W}\cos{\theta_W})$, and the above formula are in the limit of $m_\ell\ll m_F$. These three decay widths of $F$ are both scaled by $1/\theta_L^2$ and are plotted in Fig.~\ref{fig:decay-width-F}. As the decay width is proportional to the tiny $\theta_L^2$, $F^\pm$ can be a long-lived charged particle candidate in collider detection, where it can imprint peculiar tracks.

\subsection{Vector-like Lepton with Scalar (VLLWS)}\label{subsec:VF}

If we add another real scalar singlet $\phi$ on the basis of $F$, an effective Yukawa interaction is added to Lagrangian (\ref{eq:1}),
\begin{equation}\label{eq:11}
\begin{aligned}
\mathcal{L}^\phi_\text{Int} &\supset -y\phi\bar{F}^0_L \ell^0_R +{\rm h.c.}\\
&\simeq - y\phi\left(\bar{F}_L\ell_R + \bar{\ell}_R F_L+ \theta_R\bar{F} F - \theta_L \bar{\ell}\ell\right),
\end{aligned}
\end{equation}
where we only keep the lowest order of rotation angles $\theta_{L/R}$, and $y$ is the Yukawa coupling constant. There are four free parameters in this scenario
\begin{align}
\{ m_F, m_\phi, y, \theta_L \},
\end{align}
where we could omit $\theta_R$ because it is fixed by $\theta_R \approx \theta_L m_F/m_\ell$. We choose to treat $\theta_L$ as a free parameter instead of $\delta$ or $\theta_R$, because it is more directly connected to the lifetime of LLP. 
In our study, we focus on the parameter space of $m_F > m_\phi\gg m_\ell$, and $m_F>200\text{ GeV}$ is set to avoid constraints from multilepton and $Z$ boson searches, which will be discussed in Sec.~\ref{sec:excon}. We choose sizeable $y$ to make $F$ decay promptly so that $\phi$ is the only LLP in this scenario, but its value is bound to ensure safety from the measurements of LFVs, which will be discussed in Sec \ref{sec:excon}. It is possible for $\phi$ to interact with the mass term of $F$. However, since we are focusing on the parameter region where $m_F > m_\phi$, this interaction does not affect the long-lived signature of $\phi$. Therefore, we neglect this operator in the VLLWS model.

In this scenario, the primary decay width of the VLL is $\Gamma(F^\pm \to\phi \ell^\pm)$, vastly surpassing its decay to Standard Model (SM) particles, as the decay widths of the latter are suppressed by $\theta_L^2$. Consequently, we solely focus on $\Gamma(F^\pm \to\phi \ell^\pm)$ and the scalar decay channel $\Gamma(\phi\to \ell^+ \ell^-)$, as they are given by
\begin{equation}\label{eq:8}
\begin{aligned}
\Gamma(F^\pm \to\phi \ell^\pm) & =\frac{y^2(m_F^2+m_\ell^2-m_\phi^2)\sqrt{\lambda(m_F^2,m_\phi^2,m_\ell^2)}}{32\pi m_F^3},\\
\Gamma(\phi\to \ell^+ \ell^-) & = \frac{(y\theta_L)^2m_\phi(1-4m_\ell^2/m_\phi^2)^{3/2}}{8\pi},
\end{aligned}
\end{equation}
where the function of $\lambda(x,y,z)$ is given by
\begin{align}
\lambda(x,y,z)=x^2+y^2+z^2-2xy-2yz-2zx.
\end{align}
The decay width of the 
scalar is much smaller than the VLL by an extra small factor $\theta_L^2$, resulting $\phi$ to be a potential electrically natural long-lived particle candidate at the detector scale. Its lifetime can be expressed in the limit of $m_\phi \gg m_\ell$ as
\begin{equation}
    \tau(\phi)\simeq\left( \frac{2\times10^{-8}}{y\theta_L} \right)^2 \left( \frac{50~\text{GeV}}{m_\phi} \right)\text{ns}.
\end{equation}

\section{Existing Constraints}
\label{sec:excon}
The VLL $F$ can decay to leptons. Therefore, accounting for the constraints arising from multi-lepton searches at colliders is essential. Additionally, as the scalar particle couples with leptons, LFV constraints may apply if the Yukawa couplings to different generations of leptons exist. Both of these scenarios contribute to lepton $g-2$, further constraining these models. In this section, we will thoroughly explore all potential constraints on these two models.

To clarify, we use {\tt MadGraph 5} \cite{Alwall:2014hca} for the Monte Carlo simulations to calculate the cut efficiencies numerically in this and the following sections. The UFO models used in {\tt MadGraph 5} are generated via {\tt FeynRules} \cite{Alloul:2013bka}, based on the VLLS and VLLWS models described in Section II. The events were generated at the parton level only.

\subsection{Constraints from Heavy Stable Charged Particles Searches}
In the first scenario, the $F^\pm$ can be long-lived, while being constrained by the heavy stable charged particles (HSCPs) searches at LHC, using the data collected during 2012 \cite{CMS:2013czn} ($\sqrt{s}=8~{\rm TeV}$, $18.8~{\rm fb^{-1}}$) and 2016 \cite{CMS:2016ybj} ($\sqrt{s}=13~{\rm TeV}$, $12.9~{\rm fb^{-1}}$) runs. We can translate this constraint into the vector-like lepton with
\begin{align}
\sigma_{\rm prod}\times\epsilon(m_F,\theta_L)<\sigma_{\rm LHC},
\end{align}
where $\epsilon(m_F,\theta_L)$ is the cut efficiency on the specified parameter $m_F$ and $\theta_L$, $\sigma_{\rm LHC}$ is the cross-section upper limit (95\% C.L.) for Modified Drell-Yan Process~\cite{CMS:2013czn,CMS:2016ybj}  from LHC searches, and $\sigma_{\rm prod}$ is the production cross section in our model, also depending on $m_F$ and $\theta_L$. 
To determine the efficiency $\epsilon(m_F,\theta_L)$ for our analysis, we first utilize {\tt MadGraph 5} to generate Monte Carlo events. These events are then subjected to the ``tracker+TOF" approach, as outlined in Ref.~\cite{CMS:2013czn,CMS:2016ybj}, necessitating specific criteria such as $p_{\rm T} >65$ GeV and $|\eta|<2.1$ for $F^\pm$. To ensure comprehensive detector traversal by $F^\pm$, additional conditions are imposed: candidate tracks must be measured in the silicon-strip detector and matched to a reconstructed track in the muon system. Specifically, for $|\eta|<1.2$, a transverse decay length $r>7~{\rm m}$ is required, while for $|\eta|>1.2$, a longitudinal decay length $z>10~{\rm m}$ is mandated~\cite{CMS:2021yvr}. These conditions ensure the thorough exploration of detector space by $F^\pm$ particles. Finally, the efficiency $\epsilon(m_F,\theta_L)$ is computed by comparing the number of events passing the selection criteria to the total generated events, providing a quantitative measure of detection efficiency.
The constraint is shown in Fig.~\ref{fig:sensivqq} as the shaded gray region, where it takes two parts as these two HSCPs searches focus on different mass regions.

\subsection{Constraints from Multilepton Searches at LHC}

For the second scenario, the scalar $\phi$ can be produced from $F^\pm$ decay and then decay to a pair of lepton. If the $\phi$ is long-lived enough, it can escape the detector, which acts as missing transverse momentum $p_T^\text{miss}$ (energy $E_T^\text{miss}$). As there are two $\phi$ produced, there are two possibilities of long-lived particle search. 
If both $\phi$ escape the detector, it faces the constraint on the signature of opposite-sign same-flavor (OSSF) leptons pair plus $p_T^\text{miss}$ from ATLAS \cite{ATLAS:2019lff}. Here, the constraint from ATLAS \cite{ATLAS:2019lff} is adopted, which requires
\begin{align}
\sigma(pp\to F^-F^+){\rm Br}(F^\pm\to \ell^\pm\phi)\cdot \epsilon(\phi^\text{inv})<0.25~\text{fb} ~(~95\% ~\rm C.L.),
\end{align}
where $\epsilon(\phi^\text{inv})$ describes the probability of two $\phi$s escaping the detector.
In our setup, where $y\gg \theta_L$, the branching ratio ${\rm Br}(F^\pm\to \ell^\pm\phi)$ can be approximated as 1. 
To assess the cut efficiency, we utilized Monte Carlo simulations with {\tt MadGraph 5}, followed by the application of cuts on $p_T^\mu>10$ GeV, $E_T^{\rm miss}>110$ GeV, $|\eta|<2.7$, $m_{\ell_1\ell_2}>121.2$ GeV, $m_{\rm T2}>100$ GeV, and $\Delta R>{\rm min}\{10~{\rm GeV}/p_T^\mu,0.3\}$ following ATLAS~\cite{ATLAS:2019lff}, accounting for the survival efficiency of $\phi$ to escape the detector. Due to the similar slepton search category selections, we expect similar sensitivities from CMS~\cite{CMS:2020bfa}.
The constraints from this search are represented by gray shaded regions in Fig. \ref{fig:scalar-llps} and Fig.~\ref{fig:CEPC-res}. 
It is evident that, for a fixed value of $m_F$, the constraints weaken with an increasing mass of $\phi$ due to the diminishing cut efficiency. 
On the other hand, for a fixed $m_\phi$, the exclusion regions terminate at around $m_F\sim 800$ GeV. This is because the production cross-section $\sigma(pp\to F^+F^-)$ becomes smaller than the dilepton search constraints when $m_F \sim 800$ GeV.

If only one $\phi$ escapes the detector and the other $\phi$ decays inside promptly, this scenario faces the constraints from four-lepton searches at ATLAS \cite{ATLAS:2021wob}, which set a limit on
\begin{align}
\sigma(pp\to F^-F^+,F^\pm\to \ell^\pm\phi)\cdot\epsilon^\text{prompt-decay}(\phi_1)\cdot\epsilon^\text{inv}(\phi_2)\lesssim0.1~\text{fb} ~(~95\% \rm C.L.),
\end{align}
where ${\rm Br}(F^\pm\to \ell^\pm\phi)=1$ is applied.  
Apart from the kinematic cuts efficiency, the efficiency $\epsilon^\text{prompt-decay}(\phi_1)\cdot\epsilon^\text{inv}(\phi_2)$ also comprises the two geometric requirements, which ensures that one $\phi$ decays inside while the other escapes the detector, and it can be estimated as
\begin{equation*}
   \epsilon^\text{prompt-decay}(\phi_1)\cdot\epsilon^\text{inv}(\phi_2)\equiv \epsilon_\text{geo}\simeq \left(1- e^{-\frac{L_{\rm max}}{\beta\gamma \tau(\phi)}}\right) e^{-\frac{L_{\rm det}}{\beta\gamma \tau(\phi)}},
\end{equation*}
where $\beta\gamma$ is Lorentz boosted factor, $L_{\rm max}\sim \mathcal{O}(\text{mm})$ and $L_{\rm det}\sim \mathcal{O}(\text{m})$ denote the maximal allowable moving distance for the prompt decay and minimal traveling distance for the escape of $\phi$, respectively \cite{CMS:2022nty}. By numerical calculation, the maximal $\epsilon_\text{geo}^{max}$ will fall in $[0.037\%,0.0037\%]$ when $L_{\rm det}/L_{\rm max}$ is in the range $[1000,10000]$. With $\sigma(pp\to F^-F^+)\lesssim0.07~\text{pb}$ for $m_F>200$~GeV, this geometrical cut will result in the cross-section much smaller than the experimental limit. Thereby this case is free from the four-lepton searches. The conclusion also applies to the long-lived charged lepton case.

In both scenarios, LLPs have the possibility to decay inside the detector. For the first scenario, the pair produced $F^\pm$ still can decay promptly inside the detector, just need to time the efficiency of 
\begin{equation*}
    \epsilon^F_\text{prompt}\simeq \left( 1- e^{-\frac{L_{\rm max}}{\beta\gamma \tau(F)}}\right)^2.
\end{equation*}
For the second scenario, $F^\pm$ is prompt decay because of $\phi$, while $\phi$ can also decay promptly inside the detector, with efficiency 
\begin{equation*}
    \epsilon^\phi_\text{prompt}\simeq \left( 1- e^{-\frac{L_{\rm max}}{\beta\gamma \tau(\phi)}}\right)^2.
\end{equation*}
Then both two models investigated in this study need to consider the multilepton searches at LHC \cite{CMS:2019hsm,CMS:2022nty}. 
The most stringent constraint arises from the inclusive nonresonant multilepton probes of the new singlet prompt decay VLL in Ref.~\cite{CMS:2022nty}, which excluded mass range $m_F\in[125,150]$ GeV. To translate the constraint, we need to consider the above decay efficiency, which weakens the constraint. Additionally, it's worth noting that this constraint specifically applies to tauon-philic singlet vector-like leptons. Generally, for our mass range of interest where $m_F>200$ GeV, both of our models are totally safe from this constraint.

\subsection{Constraints from Lepton Flavor Violations}
In general, we can have the $3\times 3$ mass matrix for 3 flavor leptons from SM Yukawa interaction. After adding the effective mixing term $\delta \bar F^0_L \ell^0_R + {\rm h.c.} $, we can extend the mass matrix into $4\times 4$ matrix. If we take the case where the second generation lepton $\mu$ mixed with $F$ as an example, we have the $4\times 4$ mass matrix as 
\begin{equation}\label{eq:19}
\begin{aligned}
M_{\rm mass} &=
\left (\begin{array}{ccc|c}
 m_{ee} & m_{e\mu} & m_{e\tau} & 0 \\
 m_{\mu e} & m_{\mu\mu} & m_{\mu\tau} & 0 \\
 m_{\tau e} & m_{\tau\mu} & m_{\tau \tau} & 0 \\
 \hline
 0 & \delta & 0 & m_F^0
\end{array}\right).
\end{aligned}
\end{equation}
For simplicity of parameterization, we can first diagonalize the $3\times 3$ part of the SM mass by two unitary matrices $U_L^{\rm SM}$ and $U_R^{\rm SM}$, just as Eq. (\ref{eq:3}), and after the diagonalization the mass matrix can be written as
\begin{equation}
\begin{aligned}
M_{\rm mass}' &=
\left (\begin{array}{ccc|c}
 m_e' & 0 & 0 & 0 \\
 0 & m_\mu' & 0 & 0 \\
 0 & 0 & m_\tau' & 0 \\
 \hline
 \delta\cdot x_{\rm 1} & \delta\cdot x_{\rm 2} & \delta\cdot x_{\rm 3} & m_F^0
\end{array}\right),
\end{aligned}
\end{equation}
where the parameters are determined by $x_i=U_{R,i2}^{\rm SM}$.
We can further diagonalize this mass matrix with two unitary matrices $U'_L$ and $U'_R$.  In the leading order of $\theta_L$ and $\theta_R$ for each element, their explicit forms can be written as
\begin{equation}\label{eq:LFV}
\begin{aligned}
U_L'&\simeq
\left (\begin{array}{ccc|c}
 1-\frac{1}{2}r_{1a}^2\theta_L^2  & x_1x_2 r_{12}\theta_R^2 & x_1x_3 r_{13}\theta_R^2 & -x_1 r_{1a}\theta_L \\
 -x_1x_2 r_{12}\theta_R^2  & 1-\frac{1}{2}r_{2a}^2\theta_L^2 & x_2x_3 r_{23}\theta_R^2 & -x_2 r_{2a}\theta_L \\
 -x_1x_3 r_{13}\theta_R^2  & -x_2x_3 r_{23}\theta_R^2 & 1- \frac{1}{2}r_{3a}\theta_L^2 & -x_3 r_{3a}\theta_L \\
 \hline
 x_1 r_{1a}\theta_L  & x_2 r_{2a}\theta_L & x_3 r_{3a}\theta_L & 1-\frac{1}{2}\theta_L^2(r_{1a}^2x_1^2+r_{2a}^2x_2^2+r_{3a}^2x_3^2)
\end{array}\right),\\
U_R' &\simeq
\left (\begin{array}{ccc|c}
 1-\frac{1}{2}\theta_R^2  & x_1x_2 r_{12} \theta_R^2 & x_1x_3 r_{13} \theta_R^2 & -x_1 \theta_R \\
 -x_1x_2 \theta_R^2  & 1-\frac{1}{2}\theta_R^2 & x_2x_3 r_{23} \theta_R^2 & -x_2 \theta_R \\
 -x_1x_3 \theta_R^2  & -x_2x_3 \theta_R^2 & 1- \frac{1}{2}\theta_R^2 & -x_3 \theta_R \\
 \hline
 x_1 \theta_R  & x_2 \theta_R & x_3 \theta_R & 1-\frac{1}{2}\theta_R^2(x_1^2+x_2^2+x_3^2)
\end{array}\right),
\end{aligned}
\end{equation}
where $r_{i j} \equiv m_{\ell_i}/m_{\ell_j}$, $r_{i a} \equiv m_{\ell_i}/m_{\ell_a}$, $\ell_i$ refers to $e,~\mu,~\tau$ for $i=1,2,3$, respectively, and $a$ denotes the lepton flavor mixed with $F$, for example $a=2$ when $\mu$ mixed with $F$. 
This lepton masses $m_{\ell_a}$ will receive a correction of order $\mathcal{O}((\delta/m_F)^2)$ as Eq. (\ref{ml-mass}). With these two unitary matrices, in the following two subsections, we will study the LFVs for the two scenarios introduced in Sec. \ref{subsec:VLLS} and \ref{subsec:VF}.

\subsubsection{LFV for VLLS}
After the unitary rotations, within the mass eigenstates of SM leptons and VLLs, the interaction parts of Lagrangian for VLLS can be represented as
\begin{equation}
\begin{aligned}\label{eq:15}
    \mathcal{L}_{\rm eff}^{\rm total}\supset~ & \mathcal{L}_{\rm eff}^{\rm LFC} + \mathcal{L}_{\rm eff}^{\rm LFV-F} + \mathcal{L}_{\rm eff}^{\rm LFV-L} + \mathcal{L}_{\rm eff}^{h}, \\
\end{aligned}
\end{equation}
where $\mathcal{L}^{\rm LFC}_{\rm eff}$ is the flavor conserving part without $\phi$ and Higgs interaction, $\mathcal{L}^{\rm LFV-F}_{\rm eff}$ represents the flavor violating interaction terms between $F$ and SM gauge bosons, contributing to the decays of $F$ and leptons $g-2$, $\mathcal{L}^{\rm LFV-L}_{\rm eff}$ stands for the flavor violating terms of SM leptons related to $Z$ boson, and $\mathcal{L}_{\rm eff}^{h}$ corresponds to the interaction terms with SM Higgs.
\begin{figure}[tb]
	\includegraphics[width=0.7\linewidth]{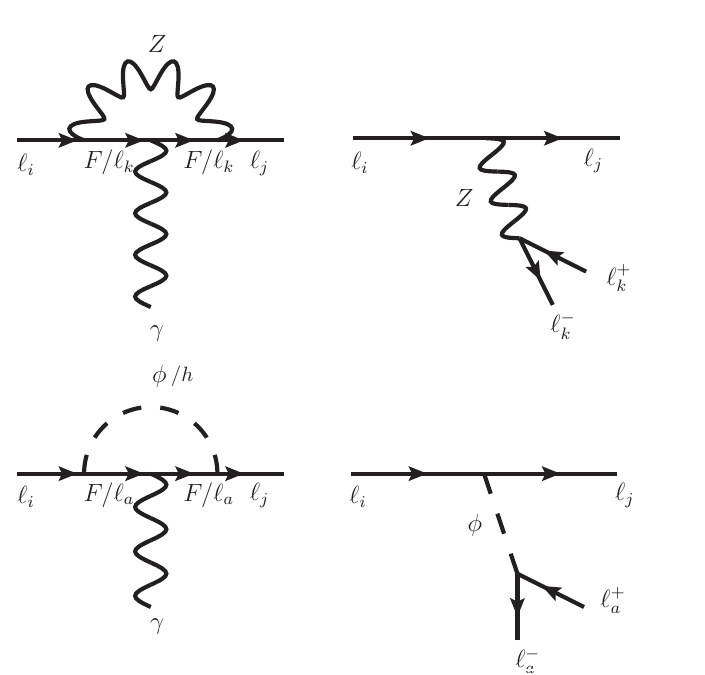}
	\centering
	\caption{Feynman diagrams for the lepton flavor violation processes.}
	\label{fig:LFV-feynman}
\end{figure}
To the leading order, the flavor-conserving part mediated by SM gauge bosons can be reduced to
\begin{equation}
\begin{aligned}\label{eq:Leff_L0}
    \mathcal{L}_{\rm eff}^{\rm LFC} =~ & \bar{F} (i\partial_\mu-e A_\mu+e\tan\theta_W Z_\mu)\gamma^\mu F - m_F\bar{F}F
    + \sum_{i=1,2,3} \bar{\ell}_i (i\partial_\mu-e A_\mu+e\tan\theta_W Z_\mu)\gamma^\mu \ell_i - m_{\ell_i}\bar{\ell}_i \ell_i \\
    & +\sum_{i=1,2,3} \frac{1}{2}\frac{e}{\sin{\theta_W}\cos{\theta_W}} Z_\mu \bar{\ell}_{iL}\gamma^\mu \ell_{iL} -\frac{e}{\sqrt{2}\sin\theta_W} (W_\mu^+ \bar{\nu}_{iL}\gamma^\mu \ell_{iL} +{\rm h.c.}),
\end{aligned}
\end{equation}
and the flavor-violating parts mediated by SM gauge bosons and the interaction terms mediated by SM Higgs can be formulated as
\begin{equation}\label{eq:24}
\begin{aligned}
    \mathcal{L}_{\rm eff}^{\rm LFV-F}&\simeq  \sum_{i=1,2,3}\left( 
    -\frac{1}{2}\frac{e}{\sin{\theta_W}\cos{\theta_W}} U_{L,44}'U_{L,i4}' Z_\mu \bar{F}_{L}\gamma^\mu \ell_{iL} -\frac{e}{\sqrt{2}\sin\theta_W} U_{L,4i}' W_\mu^+ \bar{\nu}_{iL}\gamma^\mu F_L +{\rm h.c.} \right) \\
    &\simeq \frac{e}{\sin\theta_W}\sum_{i=1,2,3} x_i r_{i a}  \theta_L \left( \frac{1}{2}\frac{1}{\cos{\theta_W}}  Z_\mu \bar{F}_{L}\gamma^\mu \ell_{iL}  - \frac{1}{\sqrt{2}} W_\mu^+ \bar{\nu}_{iL}\gamma^\mu F_L +{\rm h.c.} \right),\\
    \mathcal{L}_{\rm eff}^{\rm LFV-L} 
    &\simeq-\sum_{i,j=1,2,3,(i\neq j)} \left( \frac{1}{2}\frac{e}{\sin{\theta_W}\cos{\theta_W}} U_{L,i4}'U_{L,j4}' Z_\mu \bar{\ell}_{iL}\gamma^\mu \ell_{jL} \right)\\
    &\simeq - \frac{1}{2}\frac{e}{\sin{\theta_W}\cos{\theta_W}}\sum_{i,j=1,2,3,(i\neq j)} x_i x_j r_{i a} r_{j a} \theta_L^2  Z_\mu \bar{\ell}_{iL}\gamma^\mu \ell_{jL},\\
    &\hspace{-2.3cm}
    \mathcal{L}_{\rm eff}^{h}\simeq~ 
    \frac{h}{v_h}  (\bar{e}_L,\bar{\mu}_L,\bar{\tau}_L,\bar{F}_L)
    \left (\begin{array}{ccc|c}
 m_e & -m_e r_{2a}^2 x_1x_2 \theta_L^2 & -m_e r_{3a}^2 x_1x_3\theta_L^2 & m_e x_1 \theta_R \\
 -m_\mu r_{1a}^2 x_1 x_2\theta_L^2 & m_\mu & -m_\mu r_{3a}^2 x_2 x_3 \theta_L^2 & m_\mu x_2 \theta_R \\
 -m_\tau r_{1a}^2 x_1 x_3 \theta_L^2 & -m_\tau r_{2a}^2 x_2 x_3 \theta_L^2 & m_\tau & m_\tau x_3 \theta_R \\
 \hline
 m_e r_{1a} x_1\theta_L & m_\mu r_{2a}x_2\theta_L & m_\tau r_{3a}x_3 \theta_L  &  \left( m_e^2 x_1^2+ m_\mu^2 x_2^2+m_\tau^2x_3^2 \right)\frac{\theta_R^2}{m_F}
\end{array}\right) 
\left (\begin{array}{c}
 e_R \\
 \mu_R \\
 \tau_R  \\
 F_R \\
\end{array}\right) + {\rm h.c.}\\
&\hspace{-1.6cm}\simeq \sum_{i=1,2,3} \frac{m_{\ell_i}}{v_h}( h\bar{\ell}_{iL}\ell_{iR} + x_i\theta_R h\bar{\ell}_{iL}F_R + x_i r_{ia}\theta_L h\bar{F}_{L}\ell_{iR})
- \sum_{i,j=1,2,3,(i\neq j)} \frac{m_{\ell_i}}{v_h} x_i x_j r_{ja}^2 \theta_L^2 h\bar{\ell}_{iL} \ell_{jR}\\
&\hspace{-1cm}+\theta_R^2 \frac{\left( m_e^2 x_1^2+ m_\mu^2 x_2^2+m_\tau^2x_3^2 \right)}{v_h m_F} h \bar{F_L}F_R + {\rm h.c.}.
\end{aligned}
\end{equation}

For the electromagnetic current interaction, due to the uniform interaction among different generations, there is no electromagnetic flavor violation term. Since the new fermion $F$ is $\text{SU}(2)_L$ singlet, the LFV interactions only show up in the weak interaction through left-handed fermions mixing.
For this model, among the LFV processes related to muons, taus, and bosons ($Z$, $h$) \cite{Ardu:2022sbt}, the current strongest experimental upper limits come from the branching ratio measurement of $\mu\to eee$, $\tau\to \ell_i \ell_j \ell_j$ and $\ell_i\to \ell_j\gamma$, which respectively require \cite{Ardu:2022sbt,MEG:2016leq,SINDRUM:1987nra,BaBar:2009hkt,Hayasaka:2010np}
\begin{equation}
    {\rm Br}(\mu\to e e e)<10^{-12},~{\rm Br}(\tau\to \ell_i \ell_j \ell_j)\lesssim2.7\times10^{-8},~{\rm Br} (\ell_i\to \ell_j\gamma)\lesssim4.2\times10^{-13},
\end{equation}
where among all the LFV processes, the strongest ones for ${\rm Br}(\tau\to \ell_i \ell_j \ell_j)$ and ${\rm Br} (\ell_i\to \ell_j\gamma)$ are adopted as the upper limits (concretely, they are from the measurements of $\tau\to e\mu\mu/eee \lesssim 2.7\times10^{-8}$ and $\mu^+\to e^+\gamma \lesssim4.2\times10^{-13}$). Besides, the branching ratio is obtained by dividing by the total width of its parent particle, such as $\Gamma(\mu)=\frac{G_F^2 m_\mu^5}{192\pi^3}$, and $\Gamma(\tau)\simeq \frac{G_F^2 m_\tau^5}{192\pi^3} \cdot\frac{1}{0.178}$ with Fermi constant $G_F$.
In the upper panels of Fig.~\ref{fig:LFV-feynman}, the Feynman diagrams for these LFV processes are plotted. The corresponding branching ratios can be evaluated as
\begin{equation}\label{eq:LFV26}
\begin{aligned}
    {\rm Br}(\ell_i\to \ell_j \ell_k \ell_k)&\simeq \frac{1-4\sin^2\theta_W+8\sin^4\theta_W}{4}\cdot (x_ix_jr_{ia}r_{ja}\theta_L^2)^2\\
    &\simeq 5.0\times 10^{-11}\left( \frac{x_i}{1} \right)^2 \left( \frac{x_j}{1} \right)^2 \left( \frac{r_{ia}}{10^{-3}} \right)^2\left( \frac{r_{ja}}{1} \right)^2 \left( \frac{\theta_L}{10^{-1}} \right)^4,\\
    {\rm Br}(\ell_i\to\ell_j\gamma)&\lesssim 
    (x_ix_jr_{ia}r_{ja}\theta_L^2)^2\cdot\frac{5.6 \pi \alpha^3_{\rm EM}(11m_F^4+16 m_F^4 \sin^2 \theta_W-6m_F^2m_Z^2-33m_Z^4+36m_Z^4\ln{\frac{m_F}{m_Z}})^2}{1974799104 \sin^4\theta_W \cos^4\theta_W G_F^2m_F^8m_Z^4}\\
    &\simeq (x_ix_jr_{ia}r_{ja}\theta_L^2)^2\cdot \frac{5.6\alpha_{\rm EM}(16 \sin^2 \theta_W+11)^2}{52608\pi} \\
    &\simeq 5.3\times10^{-19}\left( \frac{x_i}{1} \right)^2 \left( \frac{x_j}{1} \right)^2 \left( \frac{r_{ia}}{10^{-3}} \right)^2\left( \frac{r_{ja}}{1} \right)^2 \left( \frac{\theta_L}{10^{-2}} \right)^4.
\end{aligned}
\end{equation}

In fact the $\mathcal{L}_{\rm eff}^h$ interaction can also contribute to the LFV $\ell_i\to\ell_j \gamma$ process just like the Feynman diagram shown in the lower left panel of Fig.~\ref{fig:LFV-feynman}. The decay width can be shown in the following:
\begin{equation}\label{eq:LFV27}
\begin{aligned}
    {\rm Br}_h(\ell_i\to\ell_j\gamma)&\simeq \left(x_ix_jr_{ia}r_{ja}\theta_L^2\frac{m_F^2}{v_h^2}\right)^2\cdot\frac{\alpha_{\rm EM}\left(m_F^6-6m_F^4m_\phi^2+3m_F^2m_\phi^4+2m_\phi^6+12m_F^2m_\phi^4\ln{\frac{m_F}{m_h}}\right)^2}{768\pi G_F^2 (m_F^2-m_h^2)^8}\\
    &< 7.8\times10^{-20}\left( \frac{x_i}{1} \right)^2 \left( \frac{x_j}{1} \right)^2 \left( \frac{r_{ia}}{10^{-3}} \right)^2\left( \frac{r_{ja}}{1} \right)^2 \left( \frac{\theta_L}{10^{-2}} \right)^4.
\end{aligned}
\end{equation}

The above calculated result shows that its contribution is much smaller than the channels mediated by $Z$ boson. At the leading order of $m_\ell$, according to the results in Ref. \cite{Ardu:2022sbt,Kuno:1999jp}, there is no interference term between the channels mediated by $Z$ boson and Higgs boson. Hence, we can directly add these two results together.
Besides, $\mathcal{L}_{\rm eff}^h$ can also contribute to the process $\ell_i\to \ell_j\ell_k\ell_k$, but compared with the contribution from $Z$ mediated process, it's obviously small according to Eq. (\ref{eq:24}) because of the additional factor $(m_i/v_h)^2$ suppression. Hence we didn't take the channel $\ell_i\to\ell_j\ell_k\ell_k$ mediated by $h$ into consideration. Finally, these constraints related to lepton decay measurement can be easily satisfied for the muon case ($a=2$), as long as we require $\theta_L<10^{-2}$ with general $x_i\sim\mathcal{O}(1)$.

\subsubsection{LFV for VLLWS}
There is the additional $\phi$-Yukawa interaction, $\mathcal{L}^{\phi}_{\rm eff}$ in the VLLWS case, which needs to be recalculated.
For this interaction, if the SM lepton in $y\phi\bar{F}_L^0 \ell_{R}^0 + {\rm h.c.}$ term is the initial flavor eigenstate, then with the unitary rotations of SM parts by $U_{L/R}^{\rm SM}$, this interaction will be transformed to
\begin{equation}\label{eq:27}
\begin{aligned}
\mathcal{L}_{\rm eff}^{\rm \phi}=~ & 
y\phi(\bar{e}_L^0,\bar{\mu}_L^0,\bar{\tau}_L^0,\bar{F}_L^0)
\left (\begin{array}{ccc|c}
 0 & 0 & 0 & 0 \\
 0 & 0 & 0 & 0 \\
 0 & 0 & 0 & 0 \\
 \hline
 0 & 1 & 0 & 0
\end{array}\right)
\left (\begin{array}{c}
 e_R^0 \\
 \mu_R^0 \\
 \tau_R^0  \\
 F_R^0 \\
\end{array}\right) + {\rm h.c.} \\
=~&y\phi(\bar{e}_L',\bar{\mu}_L',\bar{\tau}_L',\bar{F}_L')
\left (\begin{array}{ccc|c}
 0 & 0 & 0 & 0 \\
 0 & 0 & 0 & 0 \\
 0 & 0 & 0 & 0 \\
 \hline
 x_{\rm 1} &  x_{\rm 2} &  x_{\rm 3} & 0
\end{array}\right)
\left (\begin{array}{c}
 e_R' \\
 \mu_R' \\
 \tau_R'  \\
 F_R' \\
\end{array}\right) + {\rm h.c.}.
\end{aligned}
\end{equation}

After the further diagonalization of the mass matrix $M'_{\rm mass}$ via the unitary matrices $U_L'$ and $U_R'$, to the leading order, this interaction in the mass eigenstates will be finally simplified to
\begin{equation}\label{eq:28}
\begin{aligned}
\mathcal{L}_{\rm eff}^{\rm \phi}\simeq~ & 
y\phi(\bar{e}_L,\bar{\mu}_L,\bar{\tau}_L,\bar{F}_L)
\left (\begin{array}{ccc|c}
 -r_{1a}x_1^2\theta_L & -r_{1a}x_1 x_2\theta_L & -r_{1a}x_1 x_3\theta_L & 0 \\
 -r_{2a}x_1 x_2\theta_L & -r_{2a} x_2^2 \theta_L & -r_{2a}x_2 x_3\theta_L & 0 \\
 -r_{3a}x_1 x_3\theta_L & -r_{3a}x_2 x_3\theta_L & -r_{3a}x_3^2\theta_L & 0 \\
 \hline
x_1-\frac{x_1^3 \theta_R^2}{2} & x_2-\frac{x_2^3 \theta_R^2}{2} &  x_3-\frac{x_3^3 \theta_R^2}{2}  &  \theta_R \left( x_1^2+x_2^2+x_3^2 \right)
\end{array}\right)
\left (\begin{array}{c}
 e_R \\
 \mu_R \\
 \tau_R  \\
 F_R \\
\end{array}\right) + {\rm h.c.} .
\end{aligned}
\end{equation}
The last row will directly introduce order $\mathcal{O}(x_i)$ interactions between different SM lepton flavors and VLLs, mediated by scalar $\phi$, which results in sizeable contributions to the LFV processes. Especially, for $\ell_i\to \ell_j\gamma$, the branching fraction with the Feynman diagram shown in the left-lower panel of Fig. \ref{fig:LFV-feynman} can be derived as
\begin{align}
    {\rm Br}(\ell_i \to \ell_j\gamma) \simeq 
    \frac{\alpha_{\rm EM} (y^2x_ix_j)^2}{768\pi G_F^2}
    \frac{\left(m_F^6 - 6m_F^4 m_\phi^2 + 3m_F^2 m_\phi^4 + 2m_\phi^6 +
    12m_F^2 m_\phi^4 \ln\left(\frac{m_F}{m_\phi}\right)\right)^2}{\left(m_F^2 - m_\phi^2\right)^8},
\end{align}
which is derived under the condition $m_e\to 0$, $m_F>m_\phi\gg m_\mu$. The measurement of $\mu\to e\gamma$ \cite{MEG:2016leq} places strong limits on our parameters, which requires 
\begin{equation*}
    y<8.5\times10^{-3}\left(\frac{1}{x_1}\right)^{1/2}\left(\frac{1}{x_2}\right)^{1/2}\left(\frac{m_F}{200~\rm GeV}\right).
\end{equation*}

This strong constraint can be easily avoided in several cases. On the one hand, one can take the $\phi$-interaction to be under the mass eigenstate of SM leptons just as the usual considerations in $U(1)_{L_{\mu}-L_\tau}$ (the SM leptons are taken to be the mass eigenstates in the $U(1)_{L_{\mu}-L_\tau}$) \cite{He:1990pn}, which can directly prevent the mixing between different generations as Eq. (\ref{eq:27}). On the other hand, these LFVs can be well avoided or solved theoretically. The relevant terms related to LFVs come from the off-diagonal mixing between SM leptons, and this can be forbidden by introducing a global symmetry just as Ref. \cite{Bernreuther:2023uxh}, which is exactly the case we discussed in Sec. \ref{sec:model}, and it can directly prevent the mixing between different SM lepton generations. 

In our study, we follow the first case that the $\phi$-interaction is in the SM mass eigenstates, to avoid the strong LFV constraints. Then, the Lagrangian after $U'_{L/R}$ rotations can be represented as
\begin{equation}
\begin{aligned}\label{eq:Leff_phi}
\mathcal{L}^{\phi}_{\rm eff} = ~& y\phi \left(\sum_{i,j=1,2,3} \left(U_{L,i4}' U_{R,ja}' \bar{\ell}_{iL} \ell_{jR}+{\rm h.c.}\right)+\sum_{j=1,2,3} \left(U_{L,44}' U_{R,ja}' \bar{F}_{L} \ell_{jR} + U_{L,j4}' U_{R,4a}' \bar{\ell}_{jL} F_{R} + {\rm h.c.}\right) \right)\\
    &+y\phi \left( U_{L,44}' U_{R,4a}' \bar{F}_L F_R +{\rm h.c.}\right), \\
    \simeq ~& y\phi \left( x_a \theta_R \bar{F} F + (\bar{F}_L \ell_{aR} -x_a^2 \theta_L \theta_R \bar{\ell}_{aL} F_R + {\rm h.c.})
    +\sum_{i=1,2,3}\left(- x_i r_{ia}\theta_L (\bar{\ell}_{iL} \ell_{aR} + {\rm h.c.})\right)\right.  \\
    & \left.+ \sum_{i=1,2,3 (i\neq a)} \left(\mathcal{O}(x_i x_a r_{ia}^2)\cdot \theta_R^2 \bar{F}_L \ell_{iR} - x_i x_a r_{ia} \theta_L \theta_R \bar{\ell}_{iL}F_R+ {\rm h.c.} \right) + \mathcal{O}(\theta_{L/R}^3)\right).
\end{aligned}
\end{equation}
In Fig. \ref{fig:LFV-feynman}, the LFV Feynman diagrams for this model are plotted, where the lower part refers to the contributions from $\phi$-interaction. The leading contributions to the LFV processes arise mainly from the $\phi$-term, which is
\begin{equation}
\begin{aligned}
    {\rm Br}(\ell_a \to \ell_i\gamma) &\simeq 
    \frac{\alpha_{\rm EM} (x_i x_a r_{ia} y^2\theta_R^2)^2}{768\pi G_F^2}
    \frac{\left(m_F^6 - 6m_F^4 m_\phi^2 + 3m_F^2 m_\phi^4 + 2m_\phi^6 +
    12m_F^2 m_\phi^4 \ln\left(\frac{m_F}{m_\phi}\right)\right)^2}{\left(m_F^2 - m_\phi^2\right)^8}\\
    &\simeq 1.4\times10^{-15}\left( \frac{x_i}{1}\right)^2 \left( \frac{x_a}{1}\right)^2 \left( \frac{r_{ia}}{10^{-3}}\right)^2 \left( \frac{y\theta_R}{10^{-1}}\right)^4 \left( \frac{200~\rm{GeV}}{m_F}\right)^4.
\end{aligned}
\end{equation}
If we choose the parameter space as
\begin{align}
\left\{ x_i,x_a < 1, r_{ia} < 10^{-3}, y \theta_R < 10^{-1}, m_F > 200~ {\rm GeV} \right\},
\end{align}
there is no pressure from the LFV two-body decay $\ell_a \to \ell_i\gamma$ constraints.

Next, we will consider the 3-body decay process $\ell_i \to \ell_j \ell_k \ell_k$. We choose the muon-specific VLL case ($a=2$) as an example. The only kinematical allowed muon decay process is $\mu\to eee$. This LFV process can be realized by $\phi$ mediator, which is heavily suppressed by a combined factor of $\left(y^2\theta_L^2 \theta_R^2 x_1x_2r_{12} \right)^2/\left(G_F^2m_\phi^4\right)$. Thus, we only need to consider the channel mediated by $Z$ boson dominates, which is already discussed in Eq.~\eqref{eq:LFV26}. The other relevant 3-body LFV decay process is $\tau\to \mu\mu\mu$, which decay branching ratio can be estimated by 
\begin{equation}
\begin{aligned}
    {\rm Br}(\tau\to \mu \mu \mu)   \approx & \theta_L^4+\frac{\sqrt{2}(y^2x_2x_3r_{32})\theta_L^4}{16G_F m_\phi^2}+\frac{(y\theta_L)^4(x_2x_3r_{32})^2}{128G_F^2 m_\phi^4}\\
    \sim &~ 10^{-8} \left[\left(\frac{\theta_L}{10^{-2}}\right)^4 + 0.8 \left(\frac{\theta_L}{10^{-2}}\right)^2 \left(\frac{x_2}{1}\right) \left(\frac{x_3}{1}\right) \left(\frac{y\theta_L}{5\times 10^{-4}}\right)^2 \left(\frac{20~{\rm GeV}}{m_\phi}\right)^2\right.\\
    &\left.+ ~0.7\left(\frac{x_2}{1}\right)^2 \left(\frac{x_3}{1}\right)^2\left(\frac{y\theta_L}{0.5\times10^{-3}}\right)^4 \left(\frac{20~{\rm GeV}}{m_\phi}\right)^4 \right].
\end{aligned}
\end{equation}
Considering the stringent experimental constraints with $\rm Br(\tau\to \mu \mu \mu)<2.1\times 10^{-8}$  \cite{Hayasaka:2010np}, it is evident that our models are safe for this constraint, if $y\theta_L<0.5\times10^{-3}$. Therefore, combining all the LFV processes, the experimental constraints can be easily satisfied, as long as we require
\begin{equation}
    \left\{\theta_L<10^{-2},~y\theta_R<10^{-1},~y\theta_L<0.5\times10^{-3},~m_F>200~{\rm GeV~ and~} m_\phi>20~{\rm GeV}, \right\}
    \label{eq:VLLWS-oneL-masseigenstate}
\end{equation}
In addition, these conditions are consistent with the parameter regions of interest in our LLP study.

In this subsection, we have discussed lepton flavor violation in the context of the VLLWS model, which includes an additional scalar field. If the Yukawa coupling $y\phi\bar{F}_L^0 \ell_{R}^0$ involves the flavor eigenstates of the lepton, it becomes subject to strong constraints from LFV decays, such as $\mu \to e \gamma$. As a result, we are led to assume that $\ell$ must either be a SM mass eigenstate or be associated with a global symmetry that confines the coupling of $\phi$ to a single mass eigenstate. Subsequently, when considering this model, we adopt the assumption that $\ell$ is indeed a SM mass eigenstate. With this assumption in place, we can express the LFV constraints concisely through Eq.~\eqref{eq:VLLWS-oneL-masseigenstate}, which is relatively easy to meet these constraints.

\subsection{Constraints from $g-2$}

\begin{figure}[t]
\includegraphics[width=0.7\linewidth]{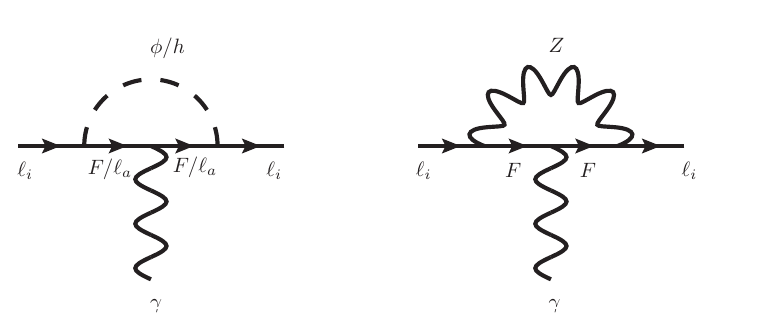}
\centering
\caption{The Feynman diagrams for the loop processes mediated by  $\phi$, $h$ and $Z$ in the context of BSM physics.}
\label{fig:g-2-feynman}
\end{figure}

The interactions as shown in the previous subsection will directly contribute to the muon and electron $g-2$. We will consider the constraints from comparing the experiment results with SM prediction \cite{Hanneke:2008tm,Morel:2020dww,Muong-2:2023cdq,Borah:2023hqw,Aoyama:2019ryr,Aoyama:2020ynm,Gnendiger:2013pva,Hoferichter:2018kwz,Colangelo:2017fiz}
\begin{equation}
\begin{aligned}
(g-2)^{\rm BSM}_\mu<2.49\times10^{-9}, \\
(g-2)^{\rm BSM}_e<9.8\times10^{-13}.
\end{aligned}
\end{equation}

In Fig.~\ref{fig:g-2-feynman}, the Feynman diagrams for the loop processes illustrate the $\phi$, $h$, and $Z$ mediated contributions in the context of BSM physics. All the contributions, to the leading order, can be  derived as
\begin{equation}\label{eq:g-2_1}
\begin{aligned}
    (g-2)_{\ell_i}^{F,\phi} &\simeq 
    \begin{dcases}
    \left(\mathcal{O}(x_i x_a r_{ia}^2)\theta_R^2(1-\delta_{ia})+\delta_{ia}\right)^2
    \frac{y^2 m_{\ell_i}^2\left(m_F^6-6m_F^4m_\phi^2+3m_F^2m_\phi^4+2m_\phi^6+12m_F^2m_\phi^4\ln{\left(\frac{m_F}{m_\phi}\right)}\right)}{48\pi^2\left(m_F^2-m_\phi^2\right)^4} \\
    &\hspace{-2cm} (m_F>m_\phi\gg m_{\ell_i}), \\
    \left(\mathcal{O}(x_i x_a r_{ia}^2)\theta_R^2(1-\delta_{ia})+\delta_{ia}\right)^2
    \frac{y^2m_{\ell_i}^2}{96\pi^2m_F^2}   &\hspace{-2cm} (m_\phi=m_F\gg m_{\ell_i}),
    \end{dcases}\\
    (g-2)_{\ell_i}^{\ell_a,\phi}&\simeq \frac{\left(y x_i r_{i a}\theta_L\right)^2 m_{\ell_i}^2}{24\pi^2 m_\phi^2},\\
    (g-2)_{\ell_i}^{F,Z}&\simeq -\left(\frac{1}{2}\frac{e}{\sin{\theta_W}\cos{\theta_W}} y x_i r_{i a}\theta_L\right)^2\cdot \left( \frac{5 m_{\ell_i}^2}{48\pi^2 m_Z^2}+\frac{m_{\ell_i}^2}{8\pi^2 m_F^2}\right),\\
    (g-2)_{\ell_i}^{F,h} &\simeq 
    \frac{m_{\ell_i}^2}{v_h^2}
    \frac{x_i^2 \theta_R^2 m_{\ell_i}^2\left(m_F^6-6m_F^4m_h^2+3m_F^2m_h^4+2m_h^6+12m_F^2m_h^4\ln{\left(\frac{m_F}{m_h}\right)}\right)}{48\pi^2\left(m_F^2-m_\phi^2\right)^4},\\
    (g-2)_{\ell_i}^{\ell_a,h}&\simeq \frac{m_{\ell_i}^2}{v_h^2}\frac{\left(x_i x_a\theta_L^2\right)^2 m_{\ell_i}^2}{24\pi^2 m_h^2}
    ~~~~~~~~~~~(a\neq i),
\end{aligned}
\end{equation}
where $\delta_{ia}$ is the Kronecker delta. The first two lines are from the contribution of $\phi$, the mediator fermions can be $F$ or SM leptons. $\phi$ predominantly couples to $\bar{F}_L \ell_{aR}$ proportional to $y$, while its coupling to $\bar{\ell}_L F_R$ is heavily suppressed, as indicated by the small parameter combination $\theta_L \theta_R$ in Eq. (\ref{eq:Leff_phi}). As a result, when concerning the $g-2$ of $\ell_a$ which $F$ directly couples to, the contribution from $F$ mediator is larger than the lepton. While for the $Z$ mediated and SM Higgs mediated contributions, they are both heavily suppressed comparing to the first line when $i=a$. For our specified parameter configuration,
\begin{equation}
\left\{
    \theta_L<10^{-2}, ~m_F>200~{\rm GeV~and}~m_\phi>20~{\rm GeV}
\right\}.
\end{equation}
Hence, we predominantly focus on the first contribution when estimating the impact of the lepton $g-2$ constraint, which leads to
\begin{equation}\label{eq:g-2_2}
\begin{aligned}
    (g-2)_{\ell_i}^{F,\phi} &\simeq
    \begin{dcases}
    5.28\times10^{-10} \left(\frac{y}{1} \right)^2 \left(\frac{m_{\ell_i}}{0.1~{\rm GeV}} \right)^2 \left(\frac{200~{\rm GeV}}{m_F} \right)^2& (m_F>m_\phi\gg m_{\ell_i}), \\
    2.64\times10^{-10} \left(\frac{y}{1} \right)^2 \left(\frac{m_{\ell_i}}{0.1~{\rm GeV}} \right)^2 \left(\frac{200~{\rm GeV}}{m_F} \right)^2 & (m_\phi=m_F\gg m_{\ell_i}) ,  
    \end{dcases}
\end{aligned}
\end{equation}
where we can typically set $y=1$ to ensure compliance with this constraint for the parameter spaces of interest to us.

To summarize the findings in Section~\ref{sec:excon}, our study primarily places constraints on the properties of long-lived charged leptons and long-lived scalars. These constraints mainly stem from searches for Heavy Stable Charged Particles at LHC and the investigation of events at LHC featuring Opposite-Sign-Same-Flavor dileptons and missing energy.
In contrast, the effects of Lepton Flavor Violation measurements and the lepton anomalous magnetic moment are negligible in the specific parameter regions that have been the focus of our investigation. For the sake of simplicity and clarity in our analysis of Long-Lived Particle signatures, we will concentrate on our chosen scenario and parameter space, and consequently, we will not consider the small LFV couplings in off-diagonal elements.

\section{Long-lived Particle Signatures at Colliders}
\label{sec:LLP}

Following the two scenarios discussed above, we will talk about the two kinds of long-lived signatures, one treats vector-like lepton $F^\pm$ as the long-lived particle, while the other treats the scalar particle $\phi$ as the long-lived particle. For both models, the VLL $F^\pm$ can be pairs produced via Drell-Yan processes at the LHC. In this section, we will explore the long-lived particle signals at the future HL-LHC (CEPC), with center-of-mass energy $\sqrt{s}=14$ TeV (240 GeV) and the integrated luminosity $\mathcal{L} = 3~\rm{ab^{-1}}$ ($5.6~\rm{ab^{-1}}$), which satisfied the constraints considered in Sec.~\ref{sec:excon}. We only consider the case where VLLs are coupled with second-generation leptons at HL-LHC to reduce the background, because due to higher misidentification of electrons, usually, electrons have higher backgrounds than muons~\cite{Reis:2016fzr, Tucker:2011zz}. While for the searches at CEPC, we assume that VLL and its accompanying scalar only mix with the first-generation lepton.
\begin{figure}[htbp]
\centering
    \includegraphics[width=0.48\linewidth]{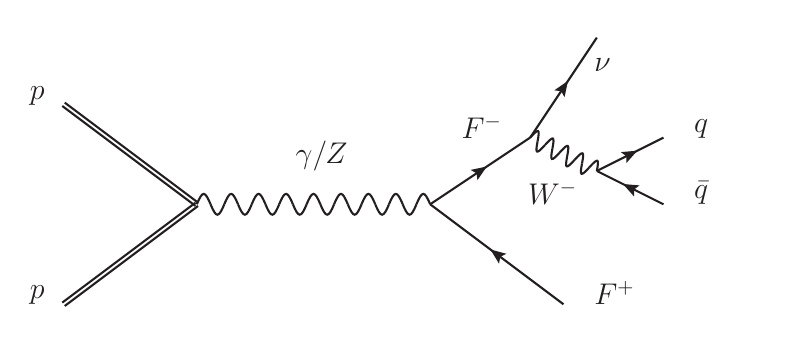}
    \includegraphics[width=0.48\linewidth]{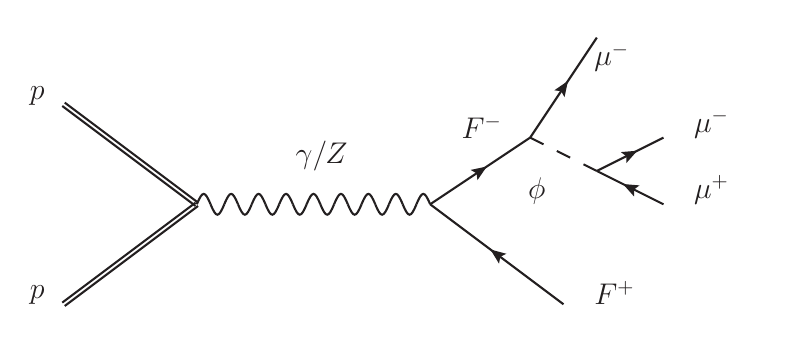}
    \caption{Feynman diagram for the production and decay of long-lived vector-like lepton and scalar at LHC. An example diagram for the signal process is shown in the left panel, and long-lived $F$ results in kink tracks signal at LHC. While in the right panel, long-lived scalar $\phi$ leads to a displaced vertex or time delay signal at LHC.}
\label{fig:prod}
\end{figure}

\subsection{Long-lived Vector-like Lepton}

In the first case, the vector-like lepton $F^\pm$ could be pair produced ($pp\to F^- F^+$) via a pure electroweak process. Then $F^\pm$ could decay into three distinct channels: $F^\pm \to W^\pm \nu$, $Z \mu^\pm$ or $h\mu^\pm$, with $\Gamma(F^\pm \to W^\pm \nu)\simeq 2 \Gamma(F^\pm \to Z \mu^\pm)\simeq 2\Gamma(F^\pm \to h\mu^\pm)$, and all widths are suppressed by the mixing angle $\theta_L$. Remarkably, all observable $F^\pm$ decay channels exhibit a comparable level of sensitivity within the parameter space of $m_F$ and $\theta_L$. Consequently, we choose the main decay $F^\pm \to W^\pm \nu$ as a representative example. We employ an inclusive search strategy that at least one of the $F^\pm$ decays inside the detector. The complete signal process can be expressed as follows:
\begin{align}\label{eq:sig1}
pp \to F^- F^+, F^- \to W^- \nu_\mu, W^- \to q\bar{q}.
\end{align}
The corresponding Feynman diagram illustrating the production and decay of this specific process is depicted in the left panel of Fig.~\ref{fig:prod}.

In order to search long-lived VLL $F^\pm$, we can effectively employ search strategies centered around kink track (KT) signatures at the HL-LHC. These strategies necessitate the reconstruction of both the charged mother and daughter particles within the tracker. To accomplish this, we must apply specific selection criteria. These distinct kink tracks can be readily identified within the Transition Radiation Tracker (TRT) component of the ATLAS detector, as described in~\cite{Asai:2011wy}. The selection conditions, following~\cite{Asai:2011wy}, are:
\begin{equation}
\label{eq:kink-track}
\begin{aligned}
    \text{KT}:~ &p_T^{F}>100~\text{GeV},~ |\eta_F|<0.63, ~0.1 < \Delta\phi <\pi/2, \\
    &~563~{\rm mm} <r_F< 863~{\rm mm},~
    |z_F|< 712~{\rm mm},\\
    &p_T^{q}>10~\text{GeV}, ~ 
    |z_q| < 712~{\rm mm~with~}
    r_q = 1066~{\rm mm},
\end{aligned}
\end{equation}
where $p_T^F$ and $p_T^{q}$ represent the transverse momentum of the parent particle $F^-$ and the daughter particle $q$ (or $\bar{q}$), respectively. $\eta_F$ is pseudorapidity of $F^-$, which cut is chosen to ensure a reasonable probability for $F^-$ to traverse the full TRT volume before decaying. $\Delta\phi$ is the kink angle (representing the relative azimuthal angle between $F^-$ and its decay products, $q$ or $\bar{q}$), which is deliberately set to a substantial value to suppress background events effectively. Additionally, $r_F$ and $|z_F|$ correspond to the transverse and longitudinal decay length of $F^-$, defined as the distance from the beam axis and along the beam axis from the nominal collision point, respectively. The cut on $r_F$ and $z_F$ ensures that the charged parent particle $F^-$ decays within the TRT volume but prior to the 3rd module. To ensure that $q$ can fly through the outermost layer of the barrel TRT, it must satisfy the condition ``$|z_q| < 712~{\rm mm~with~} r_q = 1066~{\rm mm}$''.
It is important to note that the selection conditions initially apply only to the charged leptonic decay channels of VLL. However, these conditions can be readily extended to hadronic cases. For the charged leptonic decays, the above selection conditions are good enough as the tracks of charged leptonic daughter particles can be easily reconstructed to satisfy the kink track requirement. For the hadronic case without any charged lepton, such as the signals discussed in Eq. (\ref{eq:sig1}), the selection criteria appear to require modification. However, given that the $W^\pm$ is well boosted, the partons from its decay will be confined to a narrow conical region. This situation is similar to the process $\tau'\to\tau+\nu$ with $\tau\to {\rm hadrons}$, as studied in Ref. \cite{Asai:2011wy}. Considering the prompt decay of $W^\pm$, we can treat the daughter cone as the “daughter particle” of $F^\pm$ and apply the previously mentioned selections. As a result, all decay channels of $F^\pm$ can be investigated using kink track methods, and they demonstrate similar cut efficiency.

\begin{figure}[tb]
    \centering
    \includegraphics[width=0.7\linewidth]{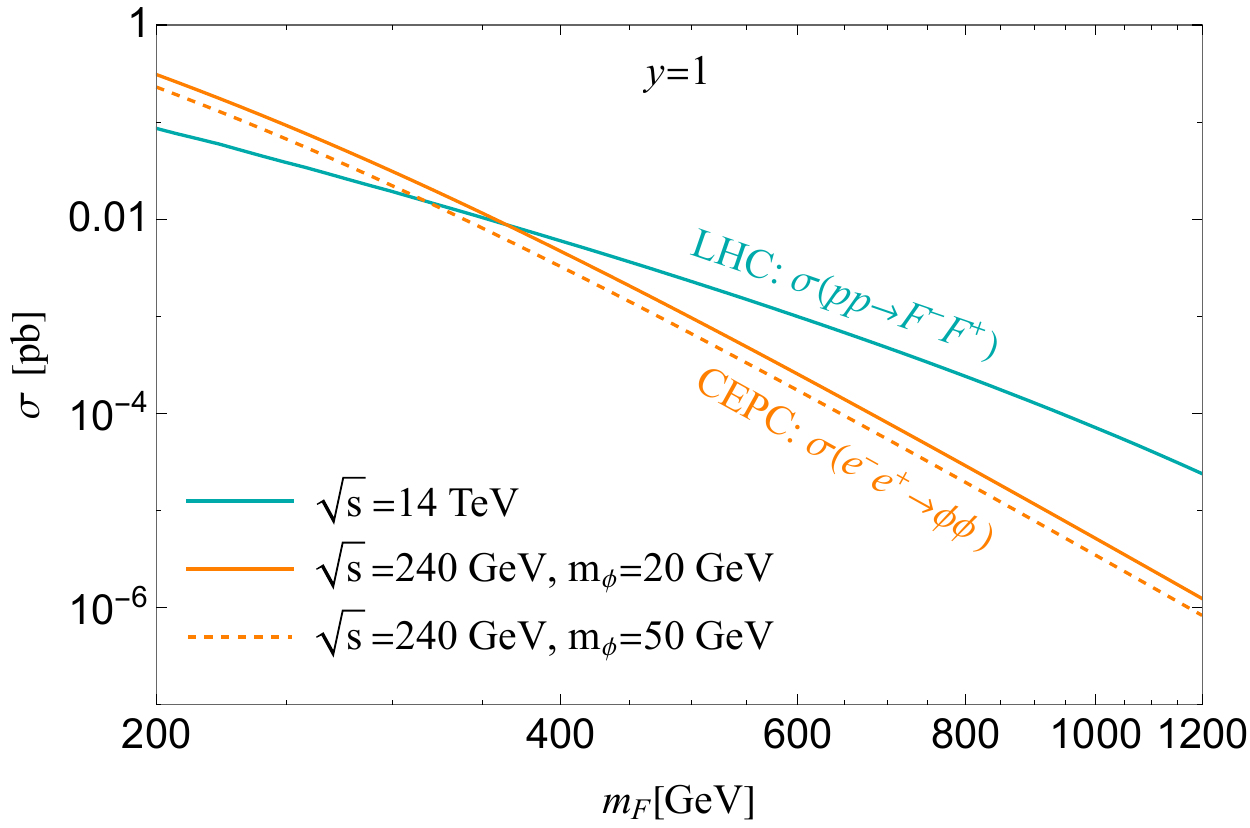}
    \caption{The production cross-sections for processes $\sigma(pp\to F^- F^+)$ of $\sqrt{s}=14$ TeV at LHC and $\sigma(e^-e^+\to \phi\phi)$ of $\sqrt{s}=240$ GeV at CEPC are shown by the cyan and the orange line, respectively. The solid orange line represents the cross-section for $m_\phi=20$ GeV, while the dashed orange line represents the cross-section for $m_\phi=50$ GeV.}
    \label{fig:xsec14TLHC}
\end{figure}

It is noteworthy that while hadronic backgrounds stemming from the SM, such as in-flight decays of $\pi^\pm$ or $K^\pm$, and stable charged hadrons undergoing hadronic interactions with trackers may modify their direction, these potential background sources may contribute to the final signal events. 
However, Ref. \cite{Asai:2011wy} argues that the background rates from three possible sources are expected to be very low at 14 TeV with an integrated luminosity of about $10~{\rm fb}^{-1}$, given the conditions: $P_T(X)>100$ GeV, $\Delta\phi>0.1$, and a reconstructed daughter track. 
Ideally, a dedicated simulation would provide a more robust verification of this argument for $3~{\rm ab}^{-1}$, which represents a 300-fold increase in integrated luminosity. However, such an analysis is beyond the scope of this work. Furthermore, kink track searches by experimental collaboration at LHC are still absent. Therefore, the zero SM background is assumed in this paper.

\begin{figure}[t]
	\includegraphics[width=0.5\linewidth]{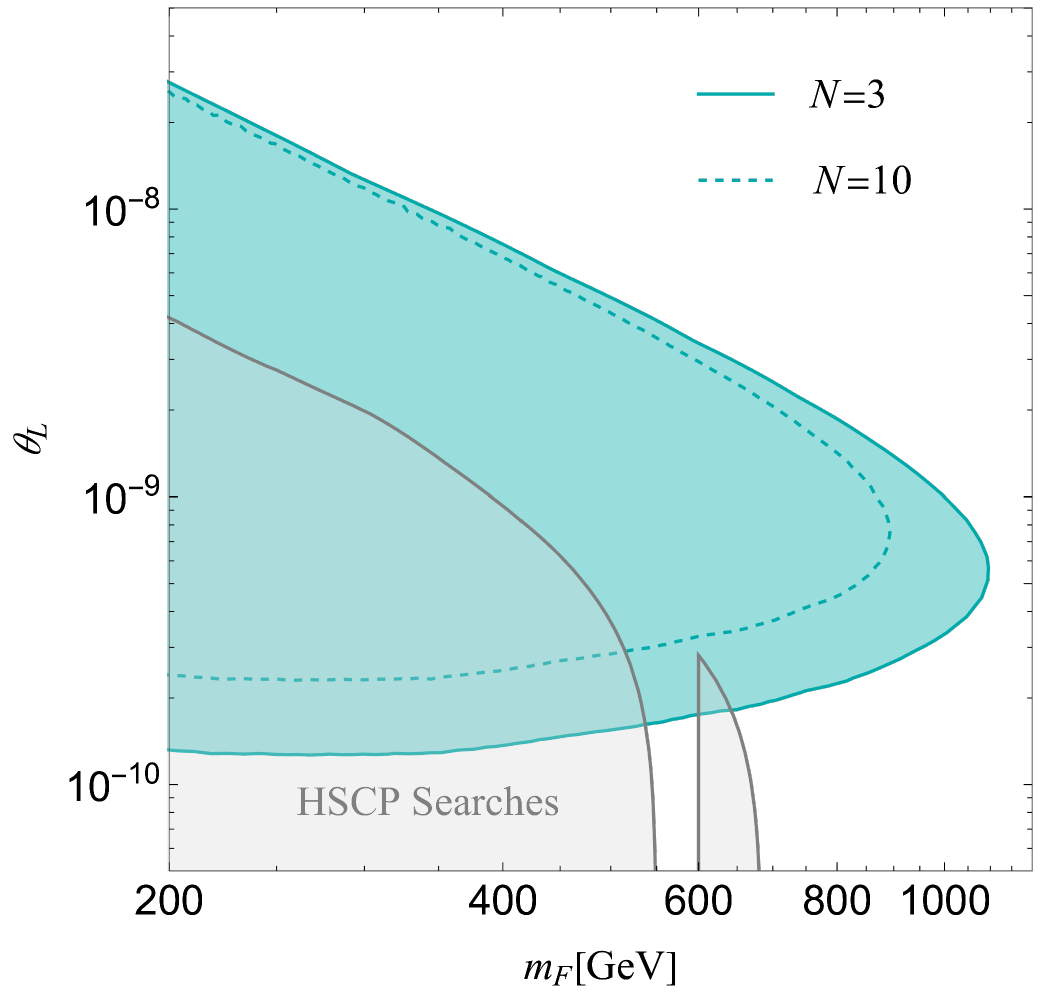}
	\centering
	\caption{The expected sensitivity for kink track signatures at HL-LHC for the long-lived vector-like lepton case is shown as a function of the vector-like lepton mass $m_F$, with integrated luminosity $\mathcal{L}=3~\text{ab}^{-1}$ and center-of-mass energy $\sqrt{s}=14$ TeV. The cyan shaded region represents a signal event number of $N_{\rm sig}\geq3$, while the dashed cyan line represents $N_{\rm sig}=10$. The gray-shaded region denotes the constraint from HSCP searches by CMS~\cite{CMS:2013czn,CMS:2016ybj}, which is divided into two parts as these two searches focus on different mass regions.
 }
	\label{fig:sensivqq}
\end{figure}

The quantity of signal events meeting these selection criteria can be determined using the following expression:
\begin{align}
N_{\rm sig} = \mathcal{L} \cdot \sigma(pp\to F^- F^+) \cdot \epsilon_{\rm cut},
\end{align}
where $\mathcal{L}=3~\text{ab}^{-1}$ represents the integrated luminosity, and $\epsilon_{\rm cut}$ accounts for the cumulative cut efficiency. The production cross-section $\sigma(pp\to F^- F^+)$ is calculated numerically utilizing {\tt MadGraph 5}, and the results are plotted in Fig.~\ref{fig:xsec14TLHC}. We ascertain the detection efficiency for the long-lived vector-like lepton with kink tracks at HL-LHC by the simulations with the {\tt MadGraph 5}. Based on the simulated cut efficiency, the expected sensitivity with all decay channels taken into consideration at HL-LHC is depicted in Fig.~\ref{fig:sensivqq}. The cyan contour region outlines the regions where the signal event $N_{\rm sig}>3$, corresponding to a $95\%$ (C.L.), while the dashed cyan line signifies $N_{\rm sig}=10$. The results illustrate that utilizing the kink track signatures can probe the parameter space for $\theta_L \in[10^{-10}, 3\times 10^{-8}]$ within the mass range of VLL $m_F \in[200, 1100]$ GeV.

\subsection{Long-lived Scalar Particle}
In the second case, the pair produced VLL $F^\pm$ promptly decays to $\phi \ell^\pm$. The scalar $\phi$ could be a long-lived particle, which acts as a displaced vertex (decay to a pair of leptons $\phi \to \ell^+ \ell^-$) or flying outside of the detector. The full signal process is
\begin{equation}
    pp\to F^-F^+, F^\pm\to\phi \ell^\pm, \phi\to \ell^+ \ell^-,
\end{equation}
and the corresponding Feynman diagram is shown in the right panel of Fig.~\ref{fig:prod}.
We consider the inclusive search, where at least one long-lived $\phi$ decays into an OSSF lepton pair inside the detector. To achieve the LLPs requirement, the lifetime and dynamical properties of $\phi$ are crucial, which depend on the $m_\phi$, $m_F$, and $\theta_L$. Especially, the dynamic property of $\phi$ and its decay products $\ell^+\ell^-$ is determined by the relationship between $m_F$ and $m_\phi$, where $\ell$ chosen as $\mu$ in this case. In this work, after assuming $y=1$, three simplified benchmarks on the free physical parameters $\{ m_F, m_\phi, \theta_L\}$ are considered, which are
\begin{equation}\label{BP-phi}
\begin{aligned}
1. \quad &\left\{m_F/m_\phi = 2,~5,~10, ~ m_F>200~{\rm GeV}, ~ \theta_L\ll1 \right\},   \\
2. \quad& \left\{m_\phi\in[20~{\rm GeV},m_F],  ~ m_F=1~{\rm TeV},~1.2~{\rm TeV}, ~ \theta_L\ll1 \right\},  \\
3. \quad & \left\{m_\phi=100~{\rm GeV},~300~{\rm GeV},~500~{\rm GeV}, ~ m_F>200~{\rm GeV}, ~ \theta_L\ll1 \right\}.
\end{aligned}
\end{equation}
Numerous strategies have been proposed to search for long-lived particles at the LHC \cite{Alimena:2021mdu}. These strategies employ distinct signatures associated with LLPs, including Displaced Vertices (DV) and Time Delays (TD), which offer effective means to reduce Standard Model (SM) backgrounds significantly. In this study, we will implement two methods characterized by these aforementioned signatures.

For both methods, it is imperative to leverage the OSSF leptons from the prompt decay of $F^\pm$. These VLLs rapidly undergo decay, resulting in a pair of high-energy and promptly detectable OSSF leptons. Such OSSF leptons can effectively serve as temporal markers for the primary vertex, triggering the identification of signal events. The concept of employing a hard lepton trigger has been explored in numerous experimental contexts. For instance, in the context of the CMS Phase-2 upgrade of the Level-1 trigger \cite{collaboration:2283192}, the trigger thresholds (incorporating a track trigger) stipulate $p_T > 27~(31)~\text{GeV}$ for isolated (non-isolated) electrons and $18~\text{GeV}$ for muons. These thresholds can be further relaxed for pairs of same-flavor leptons. As an example, within the trigger menu of ATLAS Run-2 \cite{ATL-DAQ-PUB-2019-001}, the requirements are $p_T>15~\text{GeV}~(18~\text{GeV})$ for each muon (electron), or $p_{T}^{\mu_1}>23~\text{GeV}$ and $p_{T}^{\mu_2}>9~\text{GeV}$.

Given our primary focus on heavy VLLs, whose prompt decay inherently results in two high-energy leptons, the stringent trigger threshold of $p_T^{\mu,F}>30~\text{GeV}$ is adopted in our simulations to further suppress SM backgrounds. Therefore, it is important to note that this trigger configuration is notably conservative in its approach.

\subsubsection{Displaced Vertex Search}

As discussed previously the high $p_T$ cuts on the OSSF muons pair from the vector-like $F^\pm$ decay can not only work as the trigger but also attenuate the SM backgrounds. In addition, we take into account the DV characteristics of the long-lived $\phi$ particle, which encompasses properties related to the decay position concerning the primary vertex and the displaced muon-jet (DMJ). Our selection criteria align with the specific cuts outlined in \cite{Berlin:2018jbm}:
\begin{equation}\label{eq:DMJ}
{\rm DMJ}: p_T^{\mu,F}>30~{\rm GeV},~ p_T^{\mu,\phi}>5~{\rm GeV},~ r_\phi<30~{\rm cm},~ d_0^{\mu,~\phi}>1~{\rm mm},
\end{equation}
where $p_T^{\mu,F/\phi}$ denotes the transverse momentum of muons arising from the decays of $F/\phi$, respectively. Furthermore, $r_\phi$ represents the radial displacement of the $\phi$ particle, while $d_0^{\mu,\phi}$ corresponds to the transverse impact parameter. 
It is worth noting that the application of the above cuts has been demonstrated to effectively reduce background contributions to a negligible level, as shown in Ref. \cite{Izaguirre:2015zva}. In our case, the main potential backgrounds arise from real photon conversion via collisions with material or gas in the detector, QCD hadron events, pile-up events, and tau decays. Concerning the most dominant background from real photon conversion, which has the potential to mimic the displaced lepton signal, although the production cross-section for $pp\to \mu^+\mu^-\gamma$ at $\sqrt{s}=14$ TeV with $p_T^\mu >30$ GeV can be quite large, up to $2.6$ pb, the highly suppressed photon conversion probability ($\sim m_e^2/m_\mu^2\sim 10^{-5}$) and the large invariant mass of the $\mu^+\mu^-$ pair ($m_{\mu^+\mu^-}=m_\phi>20$ GeV) result in a final effective generating cross-section for the background that is quite small and therefore negligible. As for the latter three backgrounds, similar arguments apply as in Ref. \cite{Izaguirre:2015zva}.

\subsubsection{Time Delay Search}

Time information can be effectively captured with the introduction of dedicated timing layers in future upgrades of the HL-LHC, as exemplified by CMS implementing the Minimum Ionizing Particle (MIP) timing detector \cite{CERN-LHCC-2017-027,Contardo:2020886}, ATLAS incorporating the High Granularity Timing Detector \cite{Allaire:2018bof}, and other experiments adopting similar timing layers, including LHCb \cite{LHCb:2018roe}, MATHUSLA \cite{Chou:2016lxi, Curtin:2018mvb}, FASER \cite{Kling:2018wct,Feng:2017uoz}, and CODEX-b \cite{Gligorov:2017nwh}. This invaluable timing information serves multiple purposes, as it not only addresses challenges related to pile-up events and enhances the precision of particle measurements, but also distinguishes new long-lived signals from SM backgrounds. 

Typically, SM particles travel at nearly the speed of light, while LLPs, such as $\phi$, move at much slower velocities and exhibit a noticeable time delay in their decay. The time delay can be defined as \cite{Liu:2018wte}
\begin{equation}
    \Delta t_\ell=L_\phi/\beta_\phi+L_\ell/\beta_\ell-L_{\rm{SM}}/\beta_{\rm{SM}}
\end{equation}
for this process ($\phi \to \ell \bar \ell$), where $\beta$ and $L$ denote the velocity and the moving distance, and SM denotes a trajectory from the interaction point to the arrival point at the detector by SM particle. For convenience, the trajectories of $\phi$ and decay products are assumed to be straight, and $\beta_l\simeq \beta_\text{SM} \simeq 1$.

Therefore, we also consider the TD signatures of heavy LLPs on CMS, and the cuts are chosen as follows \cite{Berlin:2018jbm,Guo:2021vpb}
\begin{equation}\label{eq:time-delay}
    {\rm TD}: p_T^{\mu,F}>30~{\rm GeV},~ p_T^{\mu,\phi}>5~{\rm GeV},~|\eta|<2.4,~ \Delta t_{\mu,\phi}>0.3~{\rm ns},~5~{\rm cm}<r_\phi<1.17~{\rm m},~ z_{\phi}<3.04~{\rm m},
\end{equation}
where $\eta$ denotes the muons pseudo-rapidity (as we choose $\mu$ for SM lepton). The cuts on the radial displacements $r_\phi$, longitudinal displacements $z_\phi$ and the TD $\Delta t_{\mu,\phi}$ ensure the decay vertex is within the CMS MIP timing detector and reduce the SM backgrounds. The simultaneous presence of promptly produced OSSF muon pairs, combined with the availability of timing information, plays a crucial role in significantly suppressing SM backgrounds, as evidenced in \cite{Liu:2018wte}. Consequently, in this study, we treat SM backgrounds as negligible. It is worth noting that several other established and proposed LHC experiments, including MATHUSLA, FASER, and CODEX-b, hold the potential to explore the long-lived signature associated with the scalar $\phi$. However, these experiments share similar features regarding constraints on the OSSF lepton pair and missing energy, as outlined in Section~\ref{sec:excon}. Consequently, we do not delve into their details within this discussion.

Based on the two search methods, the signal event number of these LLPs can be written as
\begin{equation}
    N_\text{sig}^\phi=\mathcal{L}\cdot\sigma(pp\to F^-F^+)\cdot P^\text{LLP}(\phi)\cdot \epsilon_{\rm cut},
\end{equation}
where $P^\text{LLP}(\phi)$ stands for the probability of $\phi$ decaying within the designated detector volume, and $\epsilon_{\rm cut}$ represents the other kinematic cut efficiency. To assess the signal efficiency, we conducted corresponding Monte Carlo simulations of events using {\tt MadGraph 5}. The determination of $P^\text{LLP}(\phi)$ for a given lifetime relies on the kinematics of $\phi$ and $\mu$.

\begin{figure}[tbh]
\centering
\includegraphics[width=0.32\linewidth]{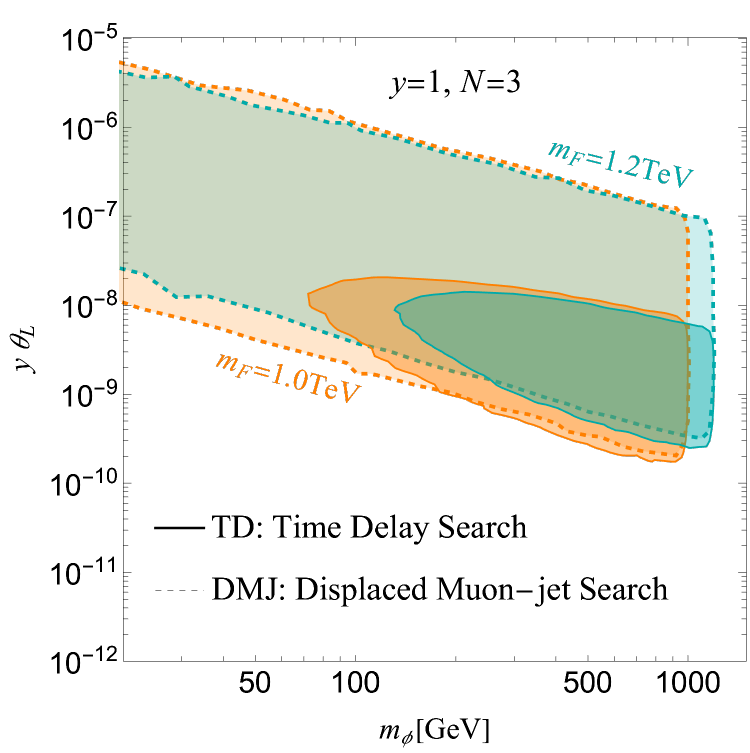}
\includegraphics[width=0.32\linewidth]{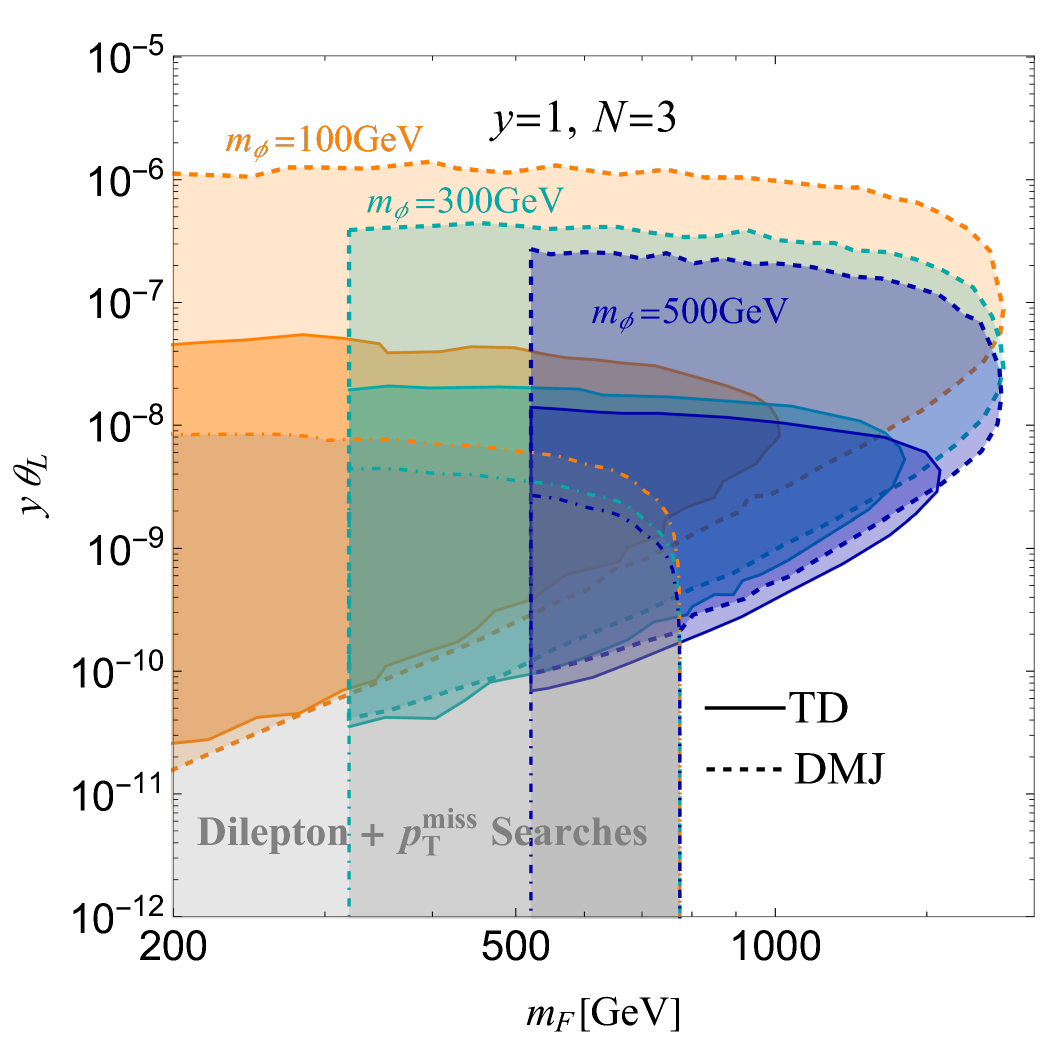}
\includegraphics[width=0.32\linewidth]{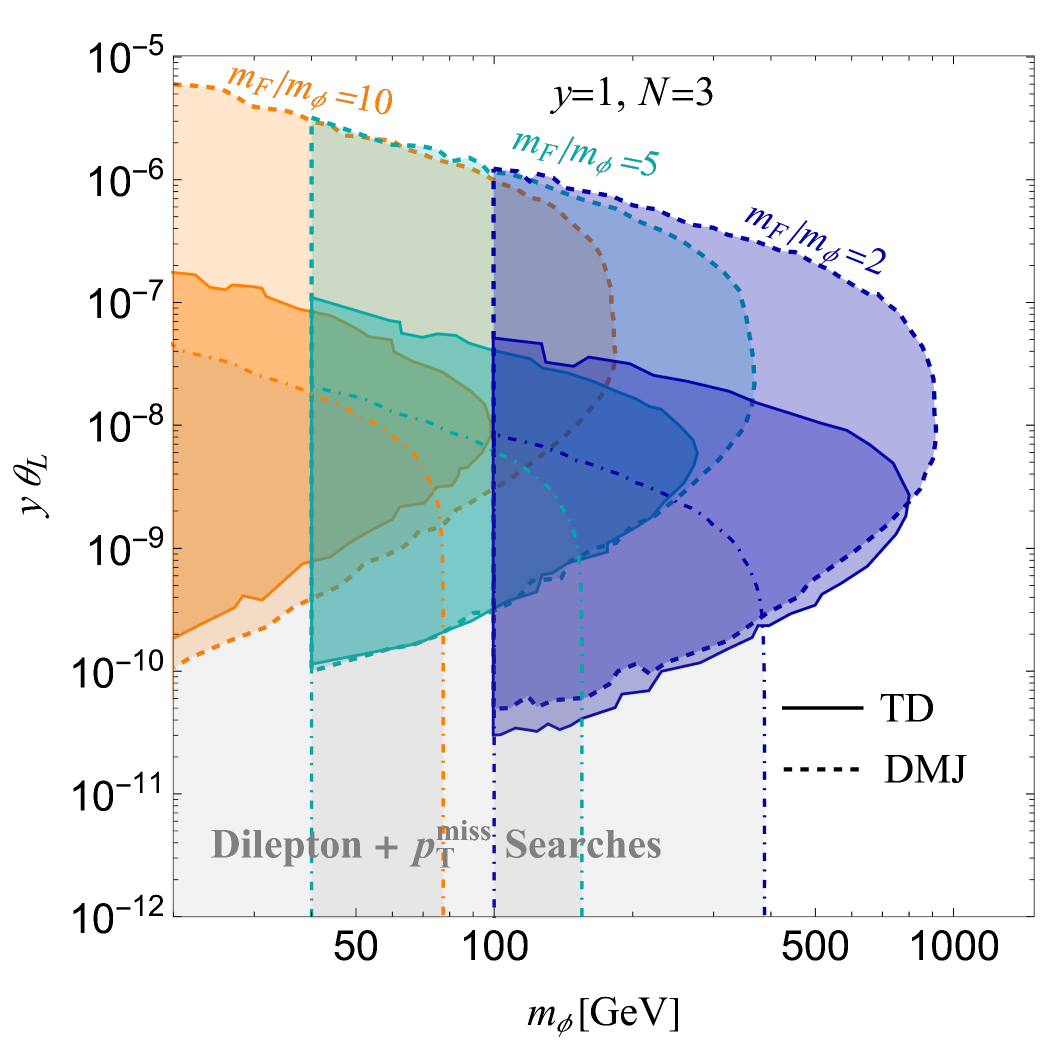}
\caption{The expected 95\% C.L. sensitivities at HL-LHC for the long-lived scalar case as a function of the scalar or vector-like lepton mass $m_\phi$ or $m_F$ for $\mathcal{L}=3~\text{ab}^{-1}$ and $\sqrt{s}=14~\text{TeV}$. From left to right, the $m_F$ fixed, $m_\phi$ fixed, and three mass ratios $m_F/m_\phi$ fixed cases are shown, respectively. Besides, the constraints from LHC dilepton and missing energy searches are also shown with gray-shaded regions with the corresponding colors as boundaries for every case.}
\label{fig:scalar-llps}
\end{figure}

Finally, the sensitivities in the $y\theta_L-m_F(m_\phi)$ plane at HL-LHC are presented in Fig.~\ref{fig:scalar-llps} for three distinct benchmarks according to Eq.~\eqref{BP-phi}. In this context, the threshold for the number of required signal events is set as $N \geqslant 3$, which precisely corresponds to the exclusion limit at a $95\%$ C.L. with zero backgrounds. A noticeable observation is that the DMJ method exhibits a preference for heavier vector-like leptons (scalar) with larger coupling strengths when compared to the TD method, primarily due to its inclination for shorter lifetimes. Additionally, as the mass of the VLL increases, the projected sensitivities weaken, primarily attributable to the marked reduction in production cross-section. 
It is important to note that the left-side or right-side truncations in each plot arise from the kinematic requirement that necessitates $m_F$ to be greater than $m_\phi$ with $m_F>200$ GeV and $m_\phi>20$ GeV. 
It is worth noting that, in contrast to the DMJ method, the TD method exhibits a closed shape for fixed $m_F$ in the left panel of Fig.~\ref{fig:scalar-llps}. This implies that the TD method lacks sensitivity in the low $\phi$ mass region. For a fixed $m_F$, when $m_F \gg m_\phi$, the $\phi$ becomes highly boosted, with $\beta_\phi \approx 1$. Consequently, its time delay becomes negligible, resulting in very low sensitivity.
Overall, these results demonstrate that the combined utilization of these two methods can achieve sensitivity in the range of $y\theta_L$ from $\sim 10^{-11}$ to $\sim 10^{-6}$ for VLL masses within the range of $m_F\in[200,1200]$ GeV. Throughout our calculations, we have taken $y=1$ as an illustrative example to ensure that the decay $F\to\phi \ell$ is both prompt and dominant, while the pair production cross-section does not depend on $y$. The other values of $y$ are also acceptable as long as the decay remains prompt and dominant.

It is important to highlight that when the mass of $\phi$ ($m_\phi$) is close to the mass of the $Z$ boson ($m_Z$), the invariant mass of the displaced or time-delayed dimuon, denoted as $m_{\mu_1\mu_2}$, will coincide with the region around the pole of the SM $Z$ boson. This scenario carries intriguing implications. If $\phi$ decays rapidly, it can potentially result in an excess of dimuon events near the $Z$ pole. This, in turn, offers a promising avenue for detecting the presence of $\phi$ with a mass in close proximity to that of the $Z$ pole by making precise measurements of the $Z$ boson.
On the other hand, one might raise concerns about whether a similar mass range could significantly affect the primary decay channel of $F^\pm$. According to Eq. (\ref{eq:FtoW}) and (\ref{eq:8}), the ratio of the two decay widths can be estimated as $\frac{\Gamma(F^-\to W^- \nu)}{\Gamma(F^-\to\phi \ell^-)}\sim \left( \frac{\theta_L}{y} \right)^2\cdot \left( \frac{m_F}{m_W} \right)^2$, assuming the negligible lepton mass. In this context, this ratio is expected to be exceedingly small, thus ensuring that the decays of $F$ remain unaffected.

\subsection{Search for Long-lived Scalars at Future CEPC}
\begin{figure}[tpb]
\centering
	\includegraphics[width=0.48\linewidth]{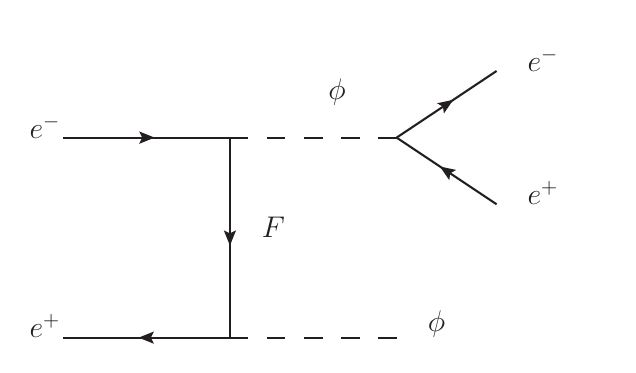}
    \includegraphics[width=0.48\linewidth]{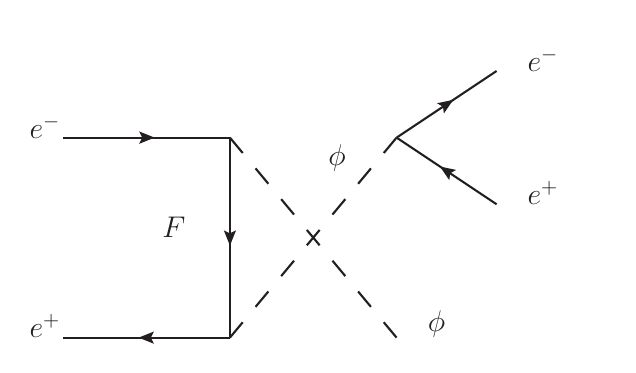}
	\caption{The Feynman diagrams for the pair production and decay of the long-lived scalar $\phi$ at CEPC, where we neglect the diagrams of exchanging electrons because of tiny coupling $\theta_L$.}
	\label{fig:cepcFD}
\end{figure}

Other than the aforementioned searches at future hadronic collider HL-LHC, the latter case of long-live scalar can also be probed at the future $e^-e^+$ collider, such as CEPC \cite{CEPC-SPPCStudyGroup:2015csa}, FCC-ee \cite{VenturiniDelsolaro:2023wqm}, and ILC \cite{ILC:2007bjz}, assuming that the $F^\pm$ and $\phi$ exclusively mix with the first SM lepton generation. Here we take CEPC with center-of-mass energies $\sqrt{s}=240$ GeV as an example. With similar parameters settings, the $F^-F^+$ pair production is forbidden due to $m_F>200$ GeV, and the only possible phenomena are the $\phi\phi$ pair production by exchanging $F^\pm$ via $t$ channel, as shown in Fig.~\ref{fig:cepcFD}. The contributions from the diagrams of exchanging electrons can be neglected, due to the suppression of tiny $\theta_L$. For the same reason, the cross-section of $\phi\gamma$ production is too small to consider.

As elucidated in Section~\ref{subsec:VF}, the scalar $\phi$ is generated and subsequently decays into a pair of $e^+e^-$ particles, leading to the manifestation of DV signatures at colliders due to its long-lived nature. Generally, the displaced vertex can be reconstructed in various detector components, including the inner tracker (IT), ECAL, HCAL, or muon spectrometer (MS), contingent upon the specific final state of its decay. In our particular scenario, featuring electronic final states ($\phi\to e^-e^+$), we exclusively employ the IT for the reconstruction of DV. This choice stems from the superior vertex reconstruction capabilities of the IT, while the ECAL and MS are less suitable for this purpose. Leveraging the efficient IT reconstruction, we implement an inclusive search strategy, stipulating the requirement for at least one reconstructed DV within the IT to confirm the identification of a signal event. The number of signal events can be expressed as:
\begin{equation}\label{eq:CEPC-signal}
    N^{\rm IT}=\mathcal{L}^{\rm CEPC}\cdot\sigma(e^-e^+\to\phi\phi)\cdot\epsilon,
\end{equation}
where $\mathcal{L}^{\rm CEPC}$ is the integrated luminosity, $\sigma(e^-e^+\to\phi\phi)$ is the cross-section of $\phi\phi$ pair production, and $\epsilon$ is the cut efficiency. The cross-section at the tree level can be formulated as
\begin{equation}
\small
    \sigma(e^-e^+\to\phi\phi)=\frac{y^2}{64 \pi s^2}\left[
    -3 \sqrt{s \left( s-4 m_\phi^2 \right)} + 
    \frac{2 \left(s^2 + 2s \left( 3m_F^2 -2m_\phi^2 \right) + 6 \left( m_F^2-m_\phi^2 \right)^2 \right) \tanh ^{-1}\frac{\sqrt{s(s-4 m_\phi^2)}}{s + 2m_F^2 - 2m_\phi^2}}{s + 2m_F^2 - 2m_\phi^2} \right],
\end{equation}
and the cross-sections with $y=1$ and $m_\phi=20$ GeV or $m_\phi=50$ GeV are plotted in Fig. \ref{fig:xsec14TLHC}.
Moreover, to calculate the efficiency, we generate Monte Carlo events by {\tt MadGraph 5} and analyze event by event. For each event, we denote the probability that at least one $\phi$ decays inside the IT as $\mathbb{P}_\phi^{\rm IT}$, where 100\% detection efficiency is assumed. The probability $\mathbb{P}_\phi^{\rm IT}$ can be expressed as \cite{Cheung:2019qdr}
\begin{equation}\label{eq:pro1}
    \mathbb{P}_\phi^{\rm IT}=\mathbb{P}_{\phi^1}^\text{IT}+\mathbb{P}_{\phi^2}^\text{IT}-\mathbb{P}_{\phi^1}^\text{IT}\cdot \mathbb{P}_{\phi^2}^\text{IT},
\end{equation}
where $\mathbb{P}_{\phi^{j}}^\text{IT}$ corresponds to the probability of the $j$th $\phi$ decaying inside the IT. Generally, for a particle $\phi^j$ produced at the interaction point (IP) and along the direction of $\Vec{r}_\phi$, the probability of it decaying within a distance $d_\phi\in [r_\phi^1, r_\phi^2]$ from the IP is given by
\begin{equation}
    \mathbb{P}_{\phi^j}^{\rm IT}=\exp\left({-\frac{r_\phi^1}{\gamma_j\beta_j\tau(\phi)}}\right)-\exp\left({-\frac{r_\phi^2}{\gamma_j\beta_j\tau(\phi)}}\right),
\end{equation}
where $\gamma_j\beta_j=\frac{E_j}{m_\phi}\frac{|\Vec{p}_j|}{E_j}$ is the Lorentz factor. In the center-of-mass frame, the two $\phi$s produced by the collisions ($e^-e^+\to\phi\phi$) hold the same magnitude momentum but opposite directions, so for axisymmetric detectors, the $\mathbb{P}_{\phi^j}^{\rm IT}$ will be same for $j=1$ and 2.
Thus, the average probability in Eq. (\ref{eq:pro1}) can be reduced to
\begin{equation}\label{eq:pro1-red}
    \mathbb{P}_\phi^{\rm IT}= 2\mathbb{P}_{\phi^1}^\text{IT}-\left(\mathbb{P}_{\phi^1}^\text{IT}\right)^2 .
\end{equation}
Specific to the IT case, we require the displaced distance $d_\phi$ along its moving direction $\Vec{r}_\phi$ to satisfy $10~\text{cm}<|d_\phi\cdot \sin\theta_i|<1.8~\text{m}$ and $|d_\phi\cdot\cos\theta_i|<2.35~\text{m}$ with $\theta_i$ represents the angle between the $\phi$ of the $i$th event and the beam line axis \cite{CEPCStudyGroup:2018ghi,Cheung:2019qdr}. There has been the choice of $5~{\rm mm} <|d_\phi|< 1.22~{\rm m}$ with the assumption of a single background event~\cite{Chrzaszcz:2020emg}, but for conservative, we choose the choice from Ref.~\cite{Cheung:2019qdr}. This condition effectively suppresses the SM backgrounds from the prompt decays of $Z$ or $H$ bosons. In addition, we apply some kinematic cuts on the $ee$ final states to further reduce the SM backgrounds. The complete selection criteria, including the displaced vertex condition, are listed as follows:
\begin{equation}\label{eq:CEPC-cuts}
\begin{aligned}
    {\rm DV-CEPC}:~~&10~{\rm cm}<|d_\phi\cdot \sin\theta_i|<1.8~{\rm m},~|d_\phi\cdot\cos\theta_i|<2.35~{\rm m},\\
    &p_T^{e_i}>30~{\rm GeV},~m_{e_1e_2}(m_\phi)>20~{\rm GeV},~\Delta R>0.01,
\end{aligned}
\end{equation}
where $p_T^{e}$ is the electron transverse momentum, and $\Delta R$ denotes the opening angle of the two electrons from the $\phi$ decays, with the corresponding dedicated requirements which can maintain good tracking spatial resolution \cite{Cheung:2019qdr}. Besides, by requiring the invariant mass $m_{e_1e_2}(m_\phi)$ of the electron pair from the displaced vertex greater than 20~GeV, the possible backgrounds from the decays of long-lived SM hadrons or mesons $(\text{SM}\to e^-e^+)$ can be attenuated. With the displaced vertex, the SM background can be eliminated to a negligible level. Furthermore, we can impose both the two $\phi$s vertices in each signal event to be displaced (2DV), which greatly reduces the only possible SM background from coincidences or misreconstructions. The probability of 2DV similar to Eq. (\ref{eq:pro1-red}) can be expressed as
\begin{equation}\label{eq:pro2}
\mathbb{P}_{\phi\phi}^{\rm IT}= \mathbb{P}_{\phi^1}^\text{IT}\cdot \mathbb{P}_{\phi^2}^\text{IT}=\left( \mathbb{P}_{\phi^1}^\text{IT}\right)^2.
\end{equation}

\begin{figure}[tbh]
\centering
\includegraphics[width=0.32\linewidth]{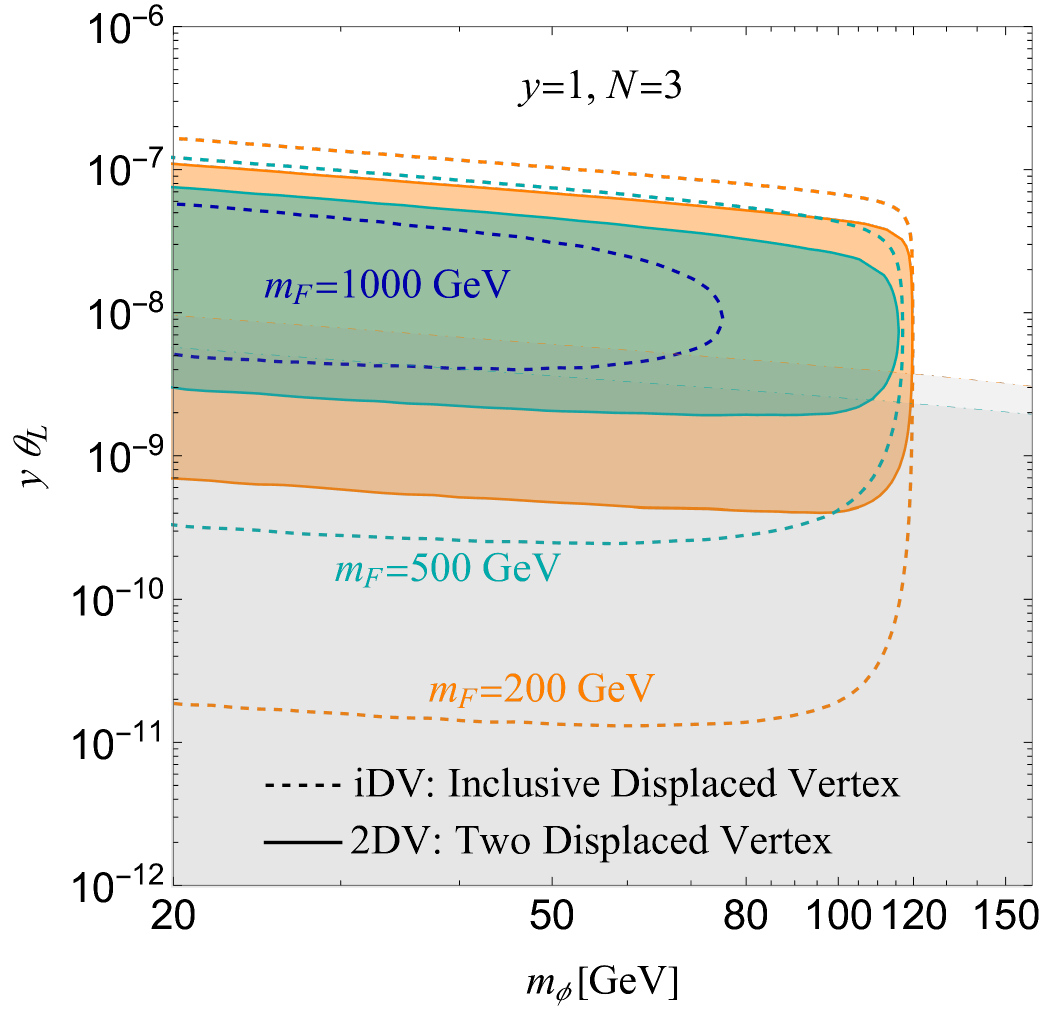}
\includegraphics[width=0.32\linewidth]{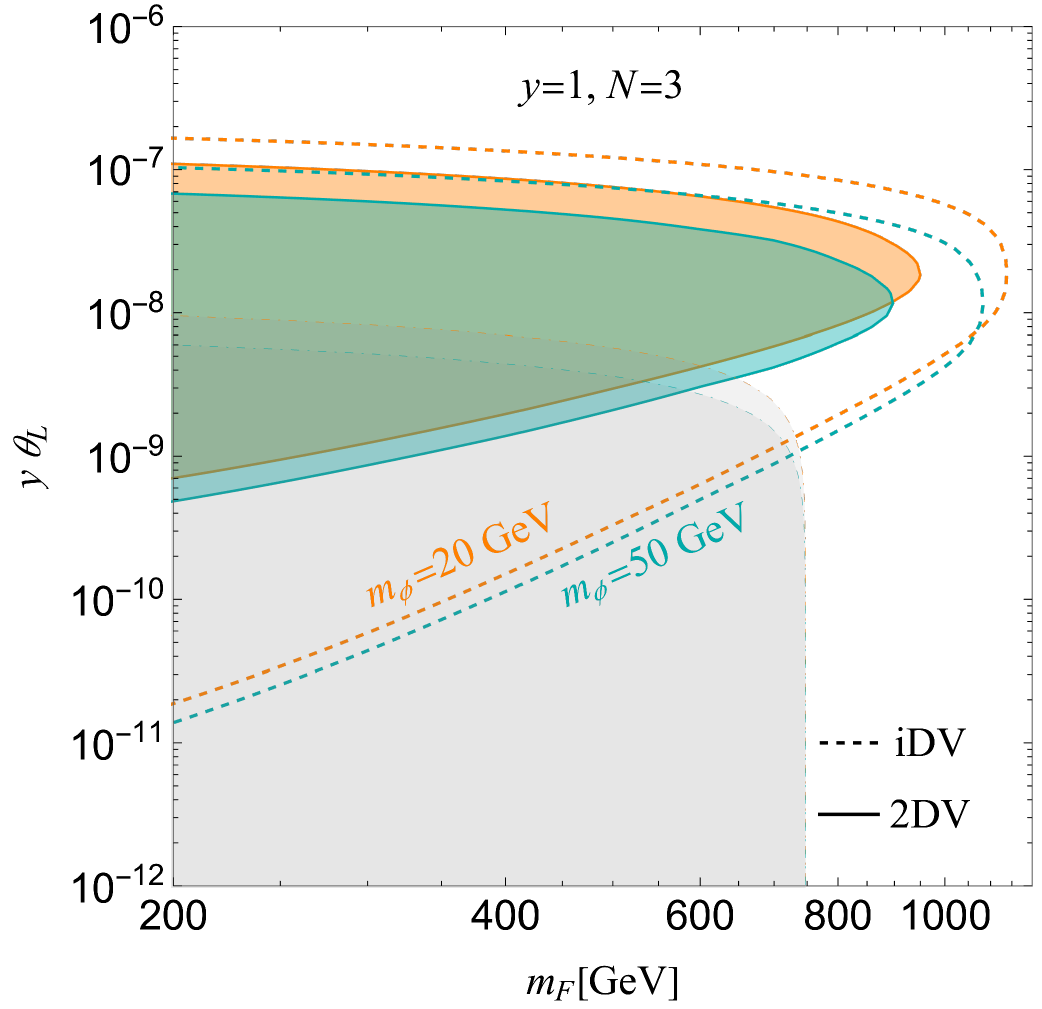}
\includegraphics[width=0.32\linewidth]{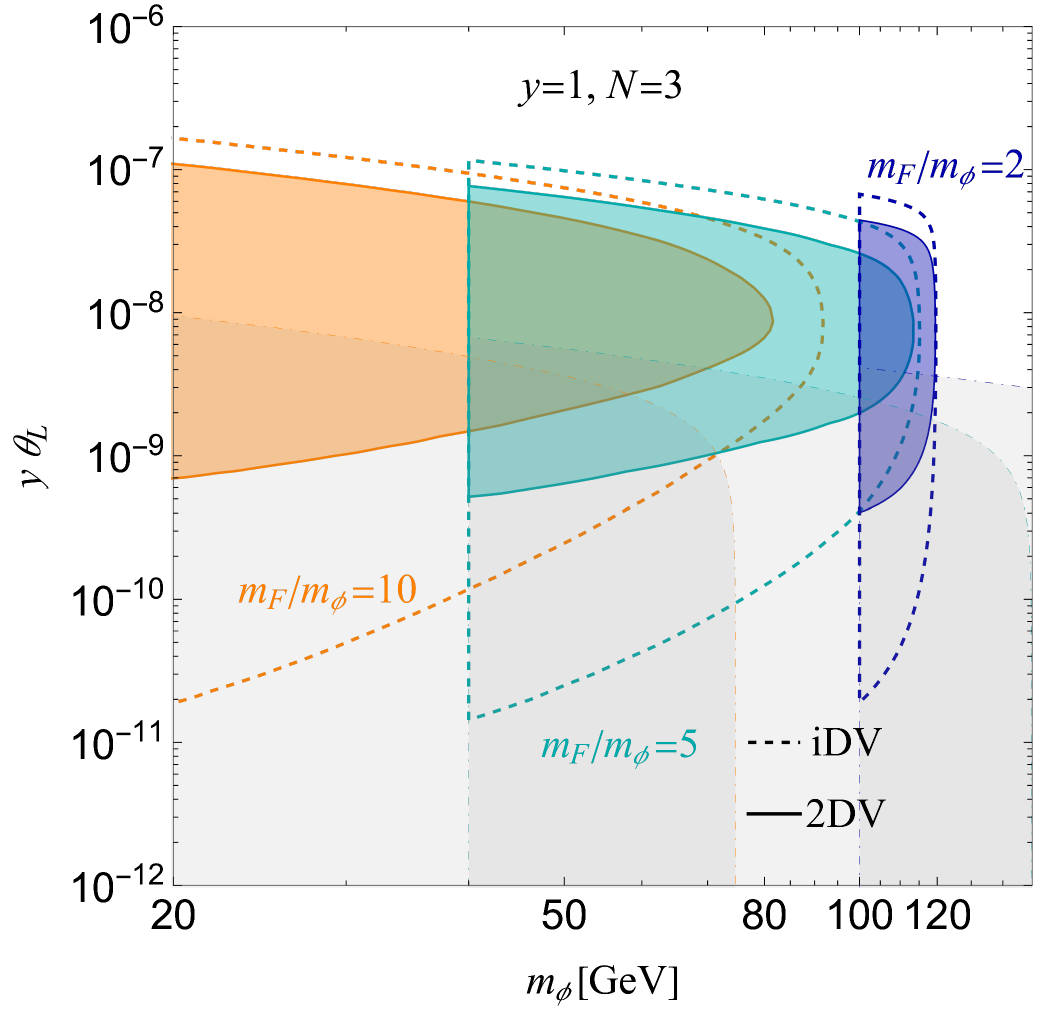}
\caption{
The expected 95\% C.L. sensitivities at Higgs factory of CEPC for the long-lived scalar case as a function of the scalar or vector-like lepton mass $m_\phi$ or $m_F$ for $\mathcal{L}=5.6~\text{ab}^{-1}$ and $\sqrt{s}=240~\text{GeV}$. From left to right, the $m_F$ fixed, $m_\phi$ fixed, and three mass ratios $m_F/m_\phi$ fixed cases are shown, respectively. The sensitivities for different masses of vector-like lepton $m_F$ are shown with different colors, and the sensitivities from iDV (inclusive displaced vertex) and 2DV (two displaced vertex) search strategies are shown with dashed lines without color shading and with straight lines with color-shading, respectively.
Moreover, the gray-shaded regions represent the constraints
from LHC dilepton and missing energy searches, with the corresponding colors as boundaries for different masses.
}
\label{fig:CEPC-res}
\end{figure}

Similar to the case of detection at LHC, we divide our parameter space $\{ m_F, m_\phi, \theta_L\}$ into three benchmarks: fixed $m_F$, fixed $m_\phi$, and fixed ratio of $m_F/m_\phi$. The corresponding sensitivities including the inclusive displaced vertex (iDV) and two displaced vertex (2DV) at future CEPC are shown in Fig. \ref{fig:CEPC-res}, with $\mathcal{L}^{\rm CEPC}=5.6~{\rm ab}^{-1}$~\cite{CEPCStudyGroup:2018ghi}, where the number of signal events $N^\text{IT}=3$ are plotted. 
The left-side or right-side truncations in each plot arise from the kinematic requirements that $\sqrt{s}>2m_\phi$ and $m_F>200$ GeV. Furthermore, the constraints on the OSSF lepton pair and missing energy are also shown in the gray-shaded regions in Fig.~\ref{fig:CEPC-res}.

Moreover, compared to HL-LHC searches, the DV searches at CEPC can well probe the low-mass regions of the scalar $\phi$ but are incapable of detecting the large mass as LHC, due to the low center-of-mass energy of CEPC. Besides, the inclusive displaced vertex searches completely cover the two displaced vertices, as expected. For $m_\phi\in[20,120]$ GeV, a small coupling $\theta_L$ of $\mathcal{O}(10^{-11},10^{-7})$ can be explored via the displaced vertex method in the future CEPC.

\section{Conclusions}
\label{sec:conclusions}

VLLs represent a straightforward extension of the SM, typically classified by their corresponding coupled SM lepton generation, implying identical quantum numbers as the SM leptons. Previous research has mainly focused on the prompt decays of the VLLs produced by electroweak processes. In this paper, we explore the potential long-lived signatures of VLLs or their subsequent decay products $\phi$, which could leave a charged long-lived track or a displaced vertex in the detector. This provides a new way of probing heavier and weakly mixed VLLs.

We consider a generic model where a new $SU(2)_L$ singlet VLL chirally mixes with its corresponding charged SM right-handed lepton via a small mass mixing term. This chiral mass mixing naturally induces small and purely left-handed current (Yukawa-like) interactions similar to the charged current, neutral current, and Higgs interactions in the SM, motivating an investigation into its long-lived signatures. 
We then introduce two specific models. In the first one, we study the straightforward searches for the long-lived charged VLLs $F^\pm$ at the ATLAS detector using the kink track method. We also examine experimental constraints from heavy stable charged particle searches, and multilepton searches, with the former excluding regions with small couplings.

Secondly, we consider another model with an additional light scalar with a sizable Yukawa interaction with the VLL. In this scenario, the long-lived signatures are transferred from the VLL to the scalar, which decays into an opposite-sign same-flavor muon pair, leaving a displaced vertex. Besides the current constraints from multilepton searches at the LHC, we also perform searches for the long-lived scalars using the displaced muon-jet and time-delay methods. These two methods show good sensitivities for $m_F\in[200,1200]$ GeV, with a moderate small coupling region around $10^{-11}<y\theta_L<10^{-6}$. Furthermore, we explore the long-lived signatures of the scalars at the future electron-positron collider CEPC, provided that the VLL couples to the first-generation SM leptons. We find that CEPC has good performance for $m_\phi<120$ GeV and $m_F<1200$ GeV, using the displaced vertex method, owing to the high luminosity of the future electron-positron collider. In conclusion, investigating the long-lived properties of VLLs or their accompanying scalars complements previous prompt searches, revealing distinctive and clean features.
 
\section{Acknowledgments}
We would like to thank Yuxuan He for his early verifying works and useful discussions. 
The work of QHC is supported in part by the National Science Foundation of China under Grants No. 11725520, No. 11675002, No. 12075257 and No. 12235001.
The work of JL is supported by Natural Science Foundation of China under grant No. 12235001 and 12075005.
The work of XPW is supported by National Science Foundation of China under Grant No. 12375095, 12005009, and the Fundamental Research Funds for the Central Universities.

\bibliographystyle{JHEP}
\bibliography{VLF_LLP.bib}
\end{document}